\documentclass[aps,pre,reprint,amsmath,amssymb]{revtex4}
\usepackage{dcolumn}
\usepackage{hyperref}
\hypersetup{
    colorlinks = true,
    citecolor = blue,
    linkcolor = red,
}

\usepackage{amsthm}
\usepackage{graphicx}
\usepackage{subcaption}
\usepackage{epsfig}
\usepackage{float}
\usepackage{scalerel}
\usepackage{dblfloatfix}
\usepackage{caption}
\usepackage{enumitem}
\usepackage[utf8]{inputenc}
\usepackage{xcolor, soul}
\sethlcolor{yellow}

\usepackage{tikz}
\newcommand{\ie}[0]{\textit{i.e.}}
\begin{document}

\preprint{APS/123-QED}
\title{Controlling the effect of quantum fluctuations in a driven nonlinear parametric oscillator}

\author{Somnath Roy}
\email{roysomnath63@gmail.com}
\affiliation{ Institute of Engineering and Management,\\ School of
University of Engineering and Management, Kolkata, 700091, India.}
\author{Chitrak Bhadra}%
\email{chitrak.iitb@gmail.com (corresponding author)}
\affiliation{Universidad Complutense de Madrid, Department of Theoretical Physics, Plaza de Ciencias, 1
Ciudad Universitaria
28040 - Madrid,Spain}
\author{Dhrubajyoti Biswas}%
\email{dhrubajyoti98@gmail.com}
\affiliation{National Brain Research Centre Gurgaon, Manesar, Haryana, 122052, India}

\begin{abstract}
This study investigates the interplay between a high-frequency external forcing and the intrinsic dynamics of a quantum nonlinear parametric oscillator.
To analyze this system, classical equations of motion of the averages of quantum operators are derived and solved by employing suitable truncation schemes and the Blekhman perturbation method. It is observed that quantum fluctuations and oscillation amplitudes within the parametric resonance zone can be modulated through the fast external periodic forcing. Moreover, the influence of the strength of driving on the overall system dynamics is systematically explored. Finally, the theoretical predictions are validated through numerical simulations, establishing the reliability of the developed framework.
\end{abstract}
\maketitle

\section{Introduction}
Parametric excitation has deep theoretical implications in the field of classical nonlinear dynamics \cite{jordan1999nonlinear,strogatz2024nonlinear} and has seen broad adoption in diverse experimental and technological domains at the nanoscale \cite{schuster2008reviews,turner_review, CPO1,Cpo3} to optical and optoelctronic platforms~\cite{optics1,optics2} and even in modelling the climate~\cite{climate_model} and the neural activities in the brain~\cite{magazu2024parametric}. Furthermore, the study of parametric oscillators, as a means to explore quantum dynamics, has garnered significant attention in modern physics due to, among other things, its importance in the motion of ions in a Paul trap \cite{RevModPhys.62.531,Paul_Trap}. Such a quantum phenomenon is at the heart of modern quantum computing, utilizing trapped ions as qubits~\cite{zoller1,zoller2}. Moreover, nanoscale devices are exploring the onset of quantum mechanical effects and thus demand our attention to study the effect of quantum fluctuations in such systems \cite{Dykman_Nano_Review_2022,Katz_Quantum_Classical,CPO2,chang2012ultrahigh,Amir_quantum_classical,teklu2015nonlinearity,Delaney_quantum_classical}.

Theoretically, the classical Mathieu oscillator, one of the simplest known models, exhibits parametric resonance due to instability. In its simplest form, the dynamics of the Mathieu oscillator is governed by a differential equation of the form
\begin{equation}
    \ddot{y}= - \omega_{0} ^{2} \left(1+ \epsilon \cos(\omega_p t)\right) y,
\label{mathieu_basic}
\end{equation}
where $y$ is the state variable representing the instantaneous displacement of the oscillator. Here, $\omega_p$ and $\epsilon$ are constants that quantify the parametric frequency and amplitude of the externally applied drive. In effect, Eq.~\eqref{mathieu_basic} describes a simple harmonic oscillator of natural frequency $\omega_0$, driven in a nonlinear fashion, and differs from a linearly driven oscillator in two aspects: (a) it exhibits a series of subharmonic resonances dictated by the condition $\omega_{0}=n\omega_p/2, (n=1,2..)$; and  (b) its amplitude grows exponentially as the oscillator phase-locks to the parametric driving while close to the resonance zones \cite{strogatz2024nonlinear, jordan1999nonlinear, turner1998five}. This is in contrast to the case of linear driving where the amplitude grows linearly.

It is well-known that the inclusion of even a small parametric driving, quantified by $\epsilon$ in Eq.~\eqref{mathieu_basic}, can lead to instabilities in the dynamics. Thus, it compels one to include nonlinear effects, which alleviates such divergences to a certain extent. To understand this intuitively, one can take into account the fact that when nonlinearity is added, the resonant frequencies of the parametric oscillator become amplitude-dependent. Thus, when large oscillations set in, the vibrational frequencies move away from the resonance condition, resulting in stable dynamics. Such a scenario can be easily investigated by adding a cubic nonlinearity or \textit{duffing nonlinearity}, resulting in a governing equation of the form
\begin{equation}
\ddot{y}= - \omega_{0} ^{2} (1+ \epsilon \cos{\omega_p t}) y - \lambda y^{3},
\label{mathieu_kerr}
\end{equation}
where $\lambda$ is a small parameter. Nonlinear detuning results from duffing nonlinearities, which are introduced by a cubic term in the restoring force. This causes the natural frequency of the oscillator to depend on its amplitude. This avoids the unbounded growth characteristic of linear Mathieu systems and permits bounded oscillations within resonance zones. Additionally, bistability, hysteresis, and more complex dynamical behaviors like chaos and nonlinear resonance are made possible by the nonlinearity \cite{ZOUNES200243,turner2,kovacic2011duffing,wang2020weak,yang2024vibrational,wiggins1987chaos}. 

The equivalent quantum problem permeates in many technological advances of modern science. From the development of nanoscale devices to one of the pioneering models of modern quantum computers, linear/nonlinear parametrically driven systems are ubiquitous in these domains as has been mentioned in some detail in the introduction. As a standard model, one can think of the quantum Hamiltonian $\hat{H}_T$, which is of the form
\begin{equation}
\hat{H}_{T}=\frac{\hat{P}^{2}}{2} + \frac{1}{2}  \omega_{0} ^{2} (1+ \epsilon \cos{\omega_p t}) \hat{X}^{2} + \sum_{k} \lambda_{k} \hat{V}_{k}[{\hat{X},t}].
\end{equation}
Here, $\hat{X}$ and $\hat{P}$ represent the position and momentum operators, whereas $\lambda_{k}$ quantifies the strengths of external static nonlinear potentials or time-dependent driving forces. For comprehensive derivations and discussions of the stability zones in the linear quantum parametric oscillator ($\lambda_{k}=0$ $\forall\ k$), the reader is directed towards some of the canonical references in the subject \cite{Paul_Trap,PhysRevA.31.564,hanggi1993dissipative,PhysRevE.52.1533}. One of the most extensive research in the theoretical aspects of the quantum nonlinear parametric oscillator with nonzero $\lambda$ (and thus especially relevant in the context of this current paper) has been carried out by Dykman et al. \cite{dykman2011quantum,zhang2017preparing,lin2015critical,marthaler2007quantum}, leading to predictions of quantum heating, quasi-energy state preparations and quantum activation among others. When linear driving is included in the above dynamics \cite{peano2014quantum}, symmetry broken spectral signatures \cite{boness2024resonant} and their utility as qubit platforms for quantum computation emerge \cite{boness2025zero}. A unifying feature in such studies is the following: the de-tuning of the parametric oscillator $|\Delta|= |\omega_p-2\omega_{0}| << \omega_{0}$, \ie, the system is resonant and weak, $\{\epsilon, |\lambda \langle \hat{X}^{2}\rangle|\} << \omega_{0}^{2}$.
In the light of such advances, one can ask the following questions:
\begin{enumerate}
\item[(a)] What is the role of quantum fluctuations in a nonlinear parametric oscillator driven by a high-frequency signal, specifically around the instability zones? 
\item[(b)] What theoretical methods are suitable to probe such quantum dynamics when the driving is at time scales much faster than any other time scales of the system?
\item[(c)] Can these stability/instability zones themselves be externally controlled/influenced via the external driving?
\end{enumerate}
It is to be noted that, throughout the paper, dissipation via a thermal bath is excluded. Furthermore, the sole focus of the current study will be centered around the \emph{Primary Subharmonic Resonance (PSR)} zone, where $\omega_p \approx 2 \omega_{0}$.

In a recent development~\cite{sarkar2020nonlinear}, approximate methods have been developed to answer the questions (a) and (b) in the autonomous case, which is the starting point of this article. The Hamiltonian studied was given by  
\begin{equation}
\hat{H}_{S}=\frac{\hat{P}^{2}}{2} + \frac{1}{2}  \omega_{0} ^{2} (1+ \epsilon \cos{\omega_p t}) \hat{X}^{2} + \frac{\lambda}{4} \hat{X}^{4} \label{Intro_JKB}.
\end{equation}
It is to be noted that, due to the quartic nonlinearity, Heisenberg's Equations of Motion (HEOM) can be utilized to derive a set of infinite hierarchy of coupled differential equations of motion of the $n$'th order average of any observable; 
for a comprehensive derivation and solutions of the hierarchy, see Ballentine et al.~\cite{PhysRevA.50.2854,ballentine1998moment,PhysRevA.65.062110}.
Approximations and truncation methods were applied to solve such essentially classical dynamical equations, using both analytical and numerical approaches. These analyses suggest non-trivial effects of quantum fluctuations on the stability/instability zones, including the existence of criticality in the fluctuation parameters.

In the current work, the Hamiltonian in Eq.~\eqref{Intro_JKB} is extended as
\begin{equation}
\hat{H}_{T}= H_{S} + \hat{U}_{\Omega}(t) \label{Intro_RBB}
\end{equation}
where an external periodic driving has been added in the form of $\hat{U}_{\Omega}(t)$, which acts on timescales of the order of $1/\Omega$. It is to be noted that the main presupposition in this work is that the external drive is assumed to be a \textit{high frequency signal} (in short, HFS) such that $\Omega\gg \textit{max} [\omega_p, \omega_{0}]$, with the region of interest being $|\Delta| \ll \omega_{0}$. This leads to a distinctly different dynamical picture from the studies where the external frequency is locked onto the parametric modulation, \ie, $\Omega =\omega_p, |\Delta|<<\omega_{0}$ \cite{peano2014quantum,boness2024resonant,boness2025zero}. With such a model in hand, the three questions posed earlier have been re-evaluated, and remarkably, a positive answer to (c) has been addressed. The results indicate that, in the presence of a sinusoidal HFS, the stability/instability zones and criticality in the fluctuations of $\hat{H}_{S}$ can be controlled externally. The response to question (a) now includes the interplay between the nonlinear coupling $\lambda$ and the HFS amplitude/period through $\hat{U}_{\Omega}(t)$ leading to corrections in all the pertinent parameters.

Furthermore, to arrive at these answers in the presence of HFS, it is realized that the theoretical methods need to be extended significantly over and above the ones pursued in Ref.~\cite{sarkar2020nonlinear}. At first, a truncation scheme leads to a closed finite set of coupled equations of motion, but they still include the HFS, acting at fast time scales, and hence a method has been resorted to that distinguishes the dynamics of a nonlinear system at widely separated timescales. The celebrated direct partition of motion introduced by Blekhman in the context of vibrational mechanics is incorporated for averaging and extracting an effective description of the driven nonlinear oscillator \cite{blekhman2000vibrational}. In effect, the method allows for a systematic treatment of the dynamics separated via two distinct timescales, and various classical nonlinear systems have been explored in this framework recently\cite{roy2020nonlinear,roy2021vibrational,roy2022hopf,ROY2023113857}. These studies, albeit classical, also point to the fact that the averaged effect of HFS can be utilised to control and modulate intrinsic system dynamics. Readers are referred to a substantial body of literature exploring the applications of high-frequency signals (HFS) in various dynamical systems, including excitable systems \cite{ullner2003vibrational}, noise-induced structures \cite{zaikin2002vibrational}, quantum models \cite{olusola2020quantum,pal2021quantum}, and multistable systems \cite{rajasekar2011novel} as well as experimental platforms\cite{chizhevsky2003experimental}. Recently, a nonlinear Langevin equation has been solved to great accuracy and opened the path for utilizing the above method in stochastic systems \cite{DSRay1}; for more details and a comprehensive review, see Ref.~\cite{yang2024vibrational}.

Since the theoretical methods can be best understood in a series of distinct steps, the rest of the paper is arranged as follows. In Sec.~\ref{sec2}, the moment equations for canonical observables and a closed, truncated set of the same involving the first two moments are derived. In Sec.~\ref{sec3}, the explicit time scale separation has been employed to derive an effective set of moment equations, which has subsumed the role of the HFS. The analytical techniques and the numerical discussions for examining the dynamics of the quantum average and the variance of the suggested model have been thoroughly addressed in Secs.~\ref{sec4} and \ref{sec5}. Finally, the article is summarized and concluded in Sec.~\ref{sec6}.

\section{The Hamiltonian and Hierarchy of the Moment Equations}
\label{sec2}
The quantum Hamiltonian that forms the foundation of this article is given by
\begin{eqnarray}
    \hat{H}_{T}=\frac{\hat{P}^{2}}{2} + \frac{1}{2}  \omega_{0} ^{2} (1+ \epsilon \cos{\omega_p t}) \hat{X}^{2} + \frac{\lambda}{4} \hat{X}^{4} + \hat{U}_{\Omega}(t).
\end{eqnarray}
The nonlinear parametric oscillator is dictated by a natural frequency $\omega_{0}$, parametric strength $\epsilon$, and parametric frequency $\omega_p$, perturbed by a quartic nonlinearity of strength $\lambda$. The system is coupled linearly to a high-frequency external drive $\hat{U}_{\Omega}(t)=G(t)\hat{X}$; here, $G(t)= -g\cos{\Omega t}$ (without any loss of generality) can be viewed as the HFS that will be utilized to control the dynamical features of $\hat{H}_{S}$. Also, $\Omega$ operates at a timescale much greater than any other dynamical scales of $\hat{H}_{S}$. For future purposes, $F(t)= 1+ \epsilon \cos(\omega_p t)$. Note, the $g=0$ case leads us back to the dynamics dictated by the Hamiltonian in Eq.~\ref{Intro_JKB}.

As a first step, the HEOM for a general operator $\hat{O}$ reads:
\begin{equation}
    \frac{d}{dt}\langle \hat{O} \rangle = \frac{\partial}{\partial t}\langle \hat{O} \rangle + \frac{1}{i \hbar} \langle[\hat{O},\hat{H}] \rangle \label{Heisenberg}
\end{equation}
\noindent The presence of nonlinearity ($\lambda$) in $\hat{H}_{T}$ leads to a hierarchy of coupled differential equations for the various quantum averages. For notational purposes the set of moments only for the variable $\hat{X}$ are described as: $\langle \hat{X} \rangle \equiv x$, $\langle (\hat{X}-x)^{2} \rangle \equiv V$ (variance), $\langle (\hat{X}-x)^{3}\rangle \equiv S$ (skewness), $\langle (\hat{X}-x)^{n} \rangle \equiv K_{n}$ (for $n>3$), etc. The hierarchy does not close, and $\langle (\hat{X}-x)^{n} \rangle$ exists for all $n$, each being coupled to a finite set of other moments depending on the exact form of the nonlinearity. A quick way to visualize the inter-relationship of the moments (denoted by arrows) for the quartic nonlinearity is given below up to the level of skewness (\ie, S). They are expressed compactly in the form of increasing time derivatives.
\begin{eqnarray}
\ddot{x} \rightarrow x, V, S;  \nonumber\\
\dddot{V} \rightarrow x, V, S, K_{1}; \nonumber\\
  \ddddot{S} \rightarrow x, V, S, K_{1} , K_{2}; .....
  \label{Hierarchy}
\end{eqnarray} 

In order to comprehend the analytical operation of the hierarchy, the first two moment equations have been explicitly expressed below (see Appendix \ref{appendix:A} for the derivation):
\begin{eqnarray}
\ddot{x} + \omega_{0}^{2} F(t)x  &=&-\lambda x^{3} - 3\lambda V x -\lambda S + g \cos{\Omega t},\nonumber\\ 
 \dddot{V} + 4\omega_{0}^{2} F(t) \dot{V} + 2 \omega_{0}^{2} \dot{F}(t) V &=&  -12 \lambda x^{2}\dot{V} - 12 \lambda V x \dot{x}  + \mathcal{F}(S,K_{1})\label{Moment0}
\end{eqnarray}
 The coupling of V to the higher moments
$S$ and $K_{1}$ are subsumed symbolically in $\mathcal{F}(S,K_{1})$ . Here, we utilize the identities $\langle \hat{X}^{3} \rangle = x^{3} + 3Vx + S$ and $\langle \hat{X}^{4} \rangle = x^{4} + 6Vx^{2} + K_{1}$. Higher moments follow similarly.
Since these equations are derived from the HEOM, it is evident how quantum effects and nonlinearity conspire together to influence the average of observables via higher-order fluctuations. Trivially, if  $ \lambda=0, \epsilon = 0, g=0$, the result of a quantum harmonic oscillator with all moments uncoupled and above $n=2$ being reducible to either 0 or powers of $V$ can be restored. It also suggests the exact solvability of the quadratic problem. The set of closed moment equations for the case $ \lambda,g=0, \epsilon \neq 0$ has been reported to represent Mathieu equations of higher-orders~\cite{hanggi1993dissipative,biswas2019properties} and has also been used to study the quantum quasi-periodic oscillator~\cite{biswas2021instability}. Different static nonlinear potentials with $\lambda \neq 0, \epsilon=0, g=0$ have been subjected to the above method recently \cite{sarkar2024quantum,pal2021quantum}.

\subsection{Hierarchical truncation and approximate closure of moment equations}
The infinite hierarchy of moment equations (\ie, Eq.~\eqref{Hierarchy}) needs to be truncated for the evaluation of any meaningful result and the footsteps of reference \cite{sarkar2020nonlinear} are being followed, by focusing on the first two moments $x(t)$ and $V(t)$ and closing the hierarchy at the level of Eq.~\eqref{Moment0}. The conditions $S=0$ and $K_{1}=3V^{2}$ have been imposed to bound the hierarchy, and their relevance will be discussed below. Proceeding with such a truncation scheme gives  (Appendix~\ref{appendix:A}),
\begin{eqnarray}
\ddot{x} + \omega_{0}^{2} F(t)x &=& -\lambda x^{3} - 3\lambda V x + g \cos{\Omega t} \nonumber\\ 
 \dddot{V} + 4\omega_{0}^{2} F(t) \dot{V} + 2 \omega_{0}^{2} \dot{F}(t) V &=&  -12 \lambda x^{2}\dot{V} - 12 \lambda V x \dot{x} - 18 \lambda V \dot{V}  \label{Moment1}
\end{eqnarray}
The dynamics of the first two quantum moments, $x(t)$ and $V(t)$, are coupled via  $\lambda$ with the hierarchy of equations closed at the second order. Under the assumptions of vanishing skewness and kurtosis, the following hold:\\
(i) an initial Gaussian wave packet $A(0) e^{-(x-x_{0})^{2}/2V_{0}}$ evolves slowly in time as  $A(t) e^{-(x-\langle x \rangle)^{2}/2V(t)}$, given initial conditions $x_{0}$ and $V_{0}$ and\\
(ii)the Gaussian character of the wavefunction is preserved over time, up to the point where the second-order moments begin to diverge.\\ Mathematically, it is evident that the extra influence of the quantum fluctuations ($V(t)$) is captured in the spread of the Gaussian wavefunction over time. More physical implications will be discussed in the subsequent sections.

 A brief but clear discussion about the above approximations is mandated. Firstly, for the purpose of rigor, it becomes evident that using the aforementioned approximation scheme to bound the hierarchy up to variance necessitates evaluating derivatives of no more than second order moments of $\hat{P}$, i.e, $\langle \hat{P} \rangle$ and $\langle(\hat{P}-\langle \hat{P}\rangle)^{2}\rangle$. However, it has been shown \cite{ballentine1998moment}, that for the explicit appearance of correction terms in non-zero powers of $\hbar$, one needs to reach at least the first derivative of $\langle(\hat{P}-\langle \hat{P}\rangle)^{3}\rangle$ for a general nonlinear potential. This explains the absence of any term involving $\hbar$ in Eq.~\eqref{Moment1} and hence it defines a semiclassical/weak quantum limit of the complete dynamics. Higher moments are notoriously difficult to evaluate and also face divergence issues~\cite{ballentine1998moment, PhysRevA.65.062110} and hence direct numerical simulation of the quantum Hamiltonian $\hat{H}_{T}$ is better for such purposes. We hope to address the full quantum problem in subsequent future works.

 Secondly, this also leads credence to the above-mentioned fact that the only variation of the Gaussian wavefunction is captured in its slow spread via a time-dependent Variance. Indeed, this is a quantum effect as the coupling of $x(t)$ to $V(t)$ only arises due to the non-zero quantum commutation relation used in deriving the hierarchy from HEOM, as well as a non-zero 
$\lambda$. 

\section{Timescale separation and the effective dynamics}
\label{sec3}
Recall that in the introduction, it is mentioned that the answer to question (b) has to involve new methods in the presence of HFS. Thus, with a set of classical nonlinear non-autonomous dynamical equations Eq.\eqref{Moment1} at hand, the effect of HFS on the evolution of the quantum averages $x(t)$ and $V(t)$ is now examined. To this end, it is noted that the HFS operates on a time scale faster than that of any other parameter in the system Hamiltonian $\hat{H}_{S}$. This leads to the utilization of time-scale separation, which is an effective description of the dynamics with the external drive being averaged out over the fast time scale \cite{blekhman2000vibrational}. The standard ansatz is
\begin{equation}
    x= s_{1}(t, \omega_{0}t) + \psi_{1}(t,\tau), \mbox{   } V= s_{2}(t, \omega_{0}t) + \psi_{2}(t, \tau), \mbox{    } \tau=\Omega t
    \label{MTA}
\end{equation}
 which decouples the dynamics in Eq.\eqref{Moment1} into distinctly separate time scales ($s_{1}/s_{2}$- slow time scale and $\psi_{1}/\psi_{2}$- fast time scale motion). Following Blekhman’s work \cite{blekhman2000vibrational}, the above form of time-scale separation in the presence of a high-frequency drive—where the fast scale will be averaged out to obtain an effective description of the dynamics—has come to be known as the Blekhman perturbation, and is now widely employed in the literature on nonlinear vibrational resonance \cite{yang2024vibrational}. A summary of the method, as relevant to the present analysis for deriving the effective dynamics, is provided in Appendix \eqref{appendix:B}. It is also noted that this marks a departure from the evaluation of the quantum dynamics governed by $\hat{H}_{T}$ (via truncated and averaged moment equations) from the case where $\omega_{p} = \Omega$ and $|\Delta| \ll \omega_{0}$, as considered in previous works \cite{peano2014quantum,boness2024resonant,boness2025zero}.

 Now, a time-averaging ($\langle .. \rangle_{f}$) is carried out where the fast time-scale dynamics is constrained by $\langle \psi_{i} \rangle_{f} = \frac{1}{2\pi} \int _{0} ^{2\pi} \psi_{i} (t, \tau) d\tau = 0$. Because of the nonlinearity, $\langle \psi_{i}^{n}\rangle_{f} (n>1)$ come into play and Eq.~\eqref{Moment1} reduces to a set of effective equations:
\begin{eqnarray}
\ddot{s}_{1} + \omega_{r}^{2} F_{r}(t)s_{1}   =- \lambda s_{1}^{3} - 3\lambda s_{1}s_{2}\nonumber\\
\dddot{s}_{2} + 4\omega_{r}^{2} F_{r}(t) \dot{s_{2}} = -2 \omega_{r}^{2}\dot{F}_{r}(t)s_{2} - 12 \lambda s_{1}^{2} \dot{s}_{2} - 18 \lambda \dot{s}_{2}s_{2} - 12 \lambda s_{2}\dot{s}_{1}s_{1}   \label{Moment2}
\end{eqnarray}
with the effective trap frequency and parametric functions given by  (see Appendix~\ref{appendix:B}) up to first order in $\epsilon$,
\begin{equation}
\omega_{r}^{2} = \omega_{0}^{2} + 3\lambda \langle \psi_{1}^{2}\rangle_{f},\mbox{  } F_{r}(t)= 1+ \epsilon_{r} \cos{\omega_{p} t}, \mbox{ 
 }\epsilon_{r} = \epsilon \omega_{0}^{2}/\omega_{r}^{2}
\end{equation}

The ratio $\omega_{0}^{2}/\omega_{r}^{2} < 1$ ensures the smallness of the parametric strength, essential to all subsequent discussions and numerics. The time-averaging on the fast time-scale has subsumed the influence of the FFT; so upto this order of perturbation, we have mapped the \textit{driven quantum oscillator} to an \textit{undriven system} (similar to the model in \cite{sarkar2020nonlinear}) \textit{with a set of redefined parameters  and controlled by the presence external driving paramterised by $g$ and $\Omega$}, ie.,
\begin{equation}
    \hat{H}_{S} + \hat{U}_{\Omega} \rightarrow \hat{H}_{S, \rm eff} =\frac{\hat{P}^{2}}{2} + \\\frac{1}{2}  \omega_{r} ^{2} (1+ \epsilon_{r} \cos{\omega_{p} t}) \hat{X}^{2} + \frac{\lambda}{4} \hat{X}^{4}
\end{equation}

Detailed calculations show that the correction term upto first order in $\epsilon$ is explicitly given by $\langle \psi_{1}^{2}\rangle_{f} = g^{2}/2 \Omega^4$; see Appendix~\ref{appendix:B}. A quick check reveals, that for $g=0$, we recover the relevant equations for $x(t)$ and $V(t)$ in reference \cite{sarkar2020nonlinear}. 
In order to study how the system evolves over the slower time scale, it is crucial to stress that the role of HFS only becomes significant when the system dynamics are averaged over the fast time scale i.e., $\tau=\Omega t$. The high-frequency signal (HFS) efficiently modulates the natural frequency \( \omega_0 \) of the system through the parameter \( \lambda g / \Omega^2 \), resulting in an effective natural frequency \( \omega_r \). As a result, this control mechanism will fail and the perturbative framework is no longer valid when nonlinearity is absent (\ie, $\lambda=0$). The behavior of the system on the slow time scale is thus specifically referred to as the ``effective description of the original dynamics", which offers a quantitative analytical understanding of the original dynamics following the averaging out of the fast-scale components. It is also noted that the method of scale separation is applied here to a coupled system of nonlinear equations, thereby illustrating the broad applicability of this approach, as previously emphasized by Blekhman \cite{blekhman2000vibrational}.

\section{Flow equations and evidence of control on dynamical features}
\label{sec4}

\begin{figure}
    \centering
    \includegraphics[scale=0.75]{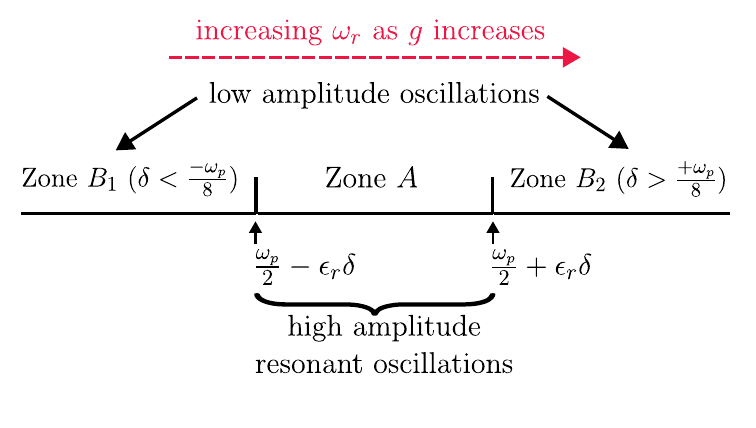}
    \caption{Representation of the three dynamical regimes based on the effective natural frequency $\omega_r$: Zone $B_1$ (low-amplitude periodic oscillations for $\omega_r < \omega_p/2 - \epsilon_r \delta$), Zone A (high-amplitude resonant oscillations within $\omega_p/2 - \epsilon_r \delta < \omega_r < \omega_p/2 + \epsilon_r \delta$), and Zone $B_2$ (low-amplitude periodic oscillations for $\omega_r > \omega_p/2 + \epsilon_r \delta$). The transitions between zones are driven by tuning the high-frequency forcing strength $g$.}

    \label{fig:schematic}
\end{figure}

Before diving into the numerical solutions of Eq.~\eqref{Moment2}, their linearized behavior around the primary subharmonic resonance (PSR) is examined. The classical equation of motion for a parametric oscillator Eq.~\eqref{mathieu_basic} reveals that the PSR zone ($\omega_{0} \approx \omega_{p}/2 + \epsilon \delta$) can be reframed as the interval $-\omega_{p}/8<\delta <\omega_{p}/8 $ (with large oscillation amplitude proportional to $\lambda^{-1/2}$)~\cite{ZOUNES200243,turner2}. This result can be directly obtained from the slow flow analysis of the equations of motion via the  Krylov-Bogoliubov Method (KBM) (for details and connection to the celebrated Rotating Wave Approximation for a similar quantum Hamiltonian, see Ref.~\cite{sarkar2020nonlinear}). In the quantum model under study, the KBM method is utilized to solve Eq.~\eqref{Moment2} around the PSR, with the newly derived effective parameters $\omega_{r} \approx \omega_{p}/2 + \epsilon_{r}\delta , \mbox{ 
     } \epsilon_{r} = \epsilon \omega_{0}^{2}/\omega_{r}^{2}, \mbox{    } \omega_{r}^{2}= \omega_{0}^{2} + (3\lambda g^{2}/2 \Omega^4)$. The KBM procedure works via the linearisation ansatz
\begin{align}
    s_{1}(t)&= A(t) \cos \left(\frac{\omega_p t}{2}\right) + B(t) \sin \left(\frac{\omega_p t}{2}\right),\\
    s_{2}(t)&= V_{0}(t) + V_{1}(t) \cos (\omega_p t) + V_{2}(t)\sin (\omega_p t).\label{KBM1}
\end{align}
For $\epsilon=0$, the solution has time-independent coefficients. Thus for small non-zero $\epsilon$, the set of amplitudes $\{A,B,V_{0}, V_{1}, V_{2}\}$ are all slowly varying functions of time and the first derivatives appear in the flow equations of the amplitude upto order $O(\epsilon)$. Eventually, a set of coupled flow equations in the five amplitude variables can be expressed as

\begin{equation}
    \begin{split}
        \dot{A}=&\epsilon_{r}\bigg(\delta-\frac{\omega_p}{8}\bigg)B+\frac{3\lambda}{4\omega_p}B(A^2+B^2)+\frac{3\lambda}{\omega_p}BV_0\\
        &+\frac{3\lambda}{2\omega_p}(AV_2-BV_1)
    \label{flow1}
    \end{split}
\end{equation}

\begin{equation}
    \begin{split}
    \dot{B}=&-\epsilon_{r}\bigg(\delta+\frac{\omega_p}{8}\bigg)A-\frac{3\lambda}{4\omega_p}A(A^2+B^2)\\
        &-\frac{3\lambda}{\omega_p}AV_0-\frac{3\lambda}{2\omega_p}(AV_1+BV_2)
        \label{flow2}
    \end{split}
\end{equation}

\begin{equation}
    \begin{split}
        \dot{V_0}=-\frac{\epsilon_r\omega_p}{4}V_2-\frac{3\lambda}{2\omega_p}(A^2-B^2)V_2+\frac{3\lambda}{\omega_p}ABV_1
        \label{flow3}
    \end{split}
\end{equation}

\begin{equation}
    \begin{split}
        \dot{V_1}&=\bigg[2\epsilon_r\delta+\frac{3\lambda}{\omega_p}(A^2+B^2)\bigg]V_2+\frac{3\lambda}{\omega_p}ABV_0\\
        &+\frac{9\lambda}{\omega_p}V_0V_2
        \label{flow4}
    \end{split}
\end{equation}

\begin{equation}
    \begin{split}
        \dot{V_2}=&-\bigg[2\epsilon_r\delta+\frac{3\lambda}{\omega_p}(A^2+B^2)\bigg]V_1\\
        &-\bigg[\frac{\epsilon_r\omega_p}{4}+\frac{3\lambda}{2\omega_p}(A^2-B^2)\bigg]V_0-\frac{9\lambda}{\omega_p}V_0V_1
        \label{flow5}
    \end{split}
\end{equation}
The above flow equations can now be used to carry out the analysis, which primarily results in the development of two zones of interest for this article: (A) the region of classical unstable PSR dictated by ($\omega_{p}/8>\delta >-\omega_{p}/8 $)  and (B) just outside the classical subharmonic resonance zone, where stable low amplitude oscillations exist ($\delta > \omega_{p}/8$ and $\delta<-\omega_{p}/8$), divided into Zone $B_1$ and $B_2$ respectively (see Fig.~\eqref{fig:schematic}).

\emph{Zone A}: At the purely classical level, Zone A reveals that the unstable solutions arise around a non-trivial fixed point ($A=0, B^{2}= \frac{4\omega_{p}}{3\lambda}[\epsilon ( \frac{\omega_{p}}{8} - \delta)])$ with no quantum fluctuations, ie., $V_{i} =0$ for $i=0,1,2$. It is accompanied by large amplitude oscillations even for small $\lambda$, as expected for a nonlinear parametric oscillator. Remarkably, the introduction of quantum fluctuations,i.e., non-zero values for $V_{i}$'s, imposes severe restrictions on such a stability picture. To see this at the analytical level, consider the case where $V_{0}$ is only a parameter in the flow equations, suppressing the time dependence in $V(t)$, \ie, $V_{1}, V_{2} \approx 0$. Immediately, it can be stated that the above nontrivial fixed point has the form
\begin{align}
A=0,\quad B^{2}= \frac{4\omega_{p}}{3\lambda}\left[\epsilon_{r} \left( \frac{\omega_{p}}{8} - \delta\right) -\frac{3\lambda V_{0}}{\omega_{p}}\right],
\end{align} 
where
\begin{align}
\epsilon_{r} &= \epsilon \omega_{0}^{2}/\left(\omega_{0}^{2} + \frac{3\lambda g^{2}}{2 \Omega^4}\right), \label{FPZoneA}
\end{align} 
which exists only if $\epsilon_{r}(\omega_{p}/8 - \delta) > 3\lambda V_{0}/\omega_{p}$ leading to the existence of a critical fluctuation parameter $V_{0c}$, which is given by
\begin{equation}
    V_{0c}=\frac{\omega_{p} \epsilon_{r}}{3\lambda}(\frac{\omega_{p}}{8} - \delta).
\end{equation}
If $V_{0}<V_{0c}$, the fixed point is a saddle, and unstable large amplitude oscillations are reported. However, when  $V_{0}>V_{0c}$, the fixed point does not exist at all (the absence of harmonic well) and trajectories become more regular with low amplitude (a similar conclusion has been addressed in Ref.~\cite{sarkar2020nonlinear} but with out the fast forcing effect, i.e. $g=0$).\\

\begin{figure}[h!]
\centering
    \subcaptionbox{$\Omega=5$.}{\includegraphics[scale=0.425]{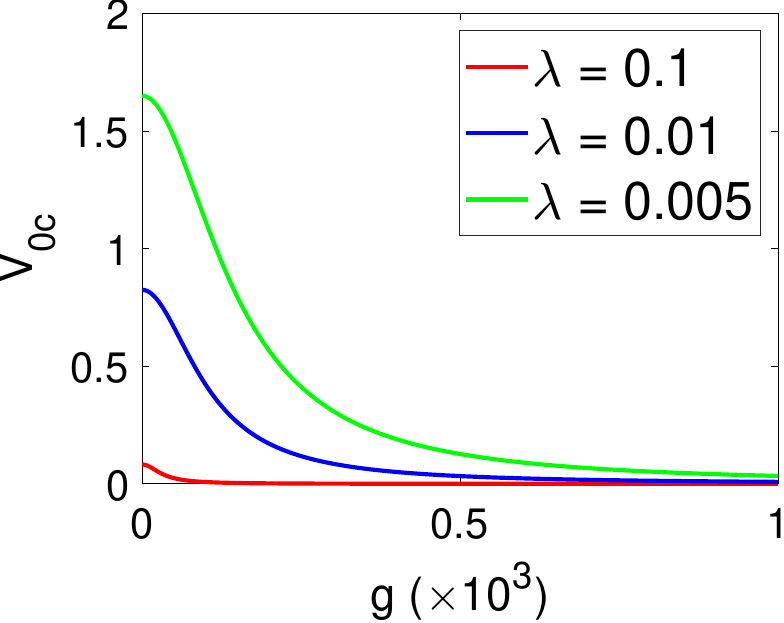}}
    \subcaptionbox{$\lambda=0.001$}{\includegraphics[scale=0.425]{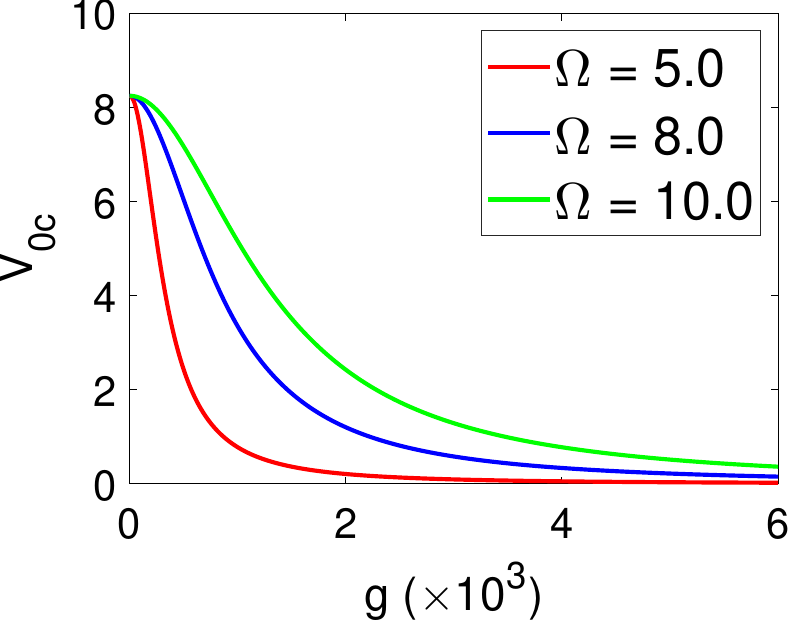}}
    \caption{Variation of $V_{0c}$ as a function of fast forcing strength $g$ for: (a) three different values of $\lambda$; and (b) for three different values of $\Omega$. The other parameters are set at $\omega_p=1.0$, $\omega_0=0.5$, and $\epsilon=0.11$; see Eq.(\ref{Vcritical}).}
    \label{fig1}
\end{figure}

The primary goal of this article is to include the impact of $g$ on this critical fluctuation, which controls the dynamics. The critical fluctuation parameter explicitly reads, as a function of $g$:
\begin{eqnarray}
      V_{0c} = \frac{\omega_{p} \epsilon}{3 \lambda}  \Big( \frac{\omega_{0}^{2}}{\omega_{0}^{2} + \frac{3 \lambda g^{2}}{2\Omega^{4}}}\Big) \Big( \frac{\omega_{p}}{8} - \delta\Big) \label{Vcritical}
\end{eqnarray}

\noindent That is, the onset of instability fixed by $V_{0c}$ can now be externally controlled via the HFS. Therefore, referring to the lesson learnt in the case of $g=0$, \textit{the erasure of the unstable oscillations} can now be controlled externally. More importantly, a glance at the expression of the fixed points in Eq.~\eqref{FPZoneA} suggests that for a given $V_0 < V_{0c}$, the amplitude of oscillation can be further suppressed by tuning the HFS strength $g$, which in turn reduces the effective parametric strength $\epsilon_r$. In fact, $g=0$ is the limit where the parametric oscillation, enslaved to the quantum fluctuation via $V_{0}$, is maximum. Based on Eq.(\ref{Vcritical}), the variation of critical fluctuation with the fast forcing strength has been depicted in Fig.\ref{fig1} to obtain the first effect of $V_{0c}$. Fig.\ref{fig1}(a) shows that when there is no external force ($g=0$), a weak nonlinear strength ($\lambda$) sets a higher critical fluctuation threshold value. 
\begin{figure*}[h!]
\centering
    \subcaptionbox{$V_0=5<V_{0c}=5.95,g=200$}{\includegraphics[scale=0.425]{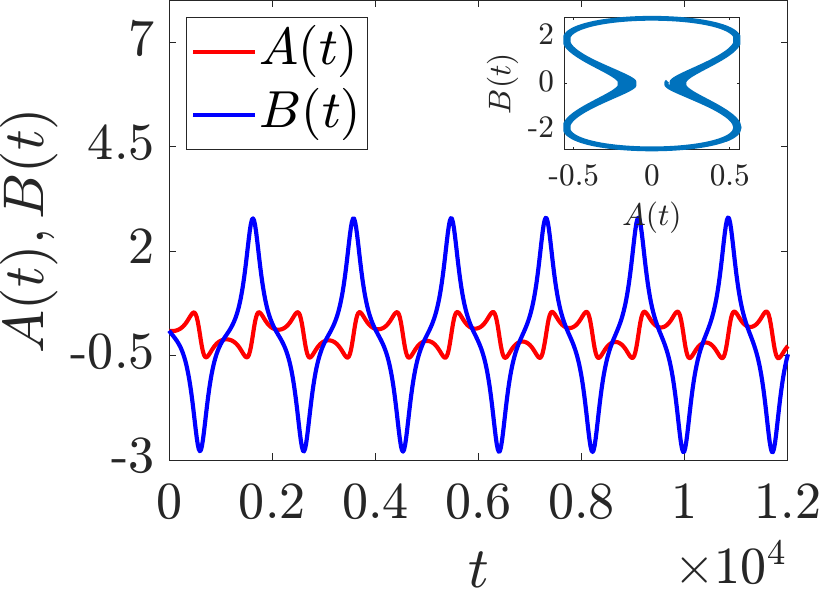}}
    \subcaptionbox{$V_0=7>V_{0c}=5.95,g=200$}{\includegraphics[scale=0.425]{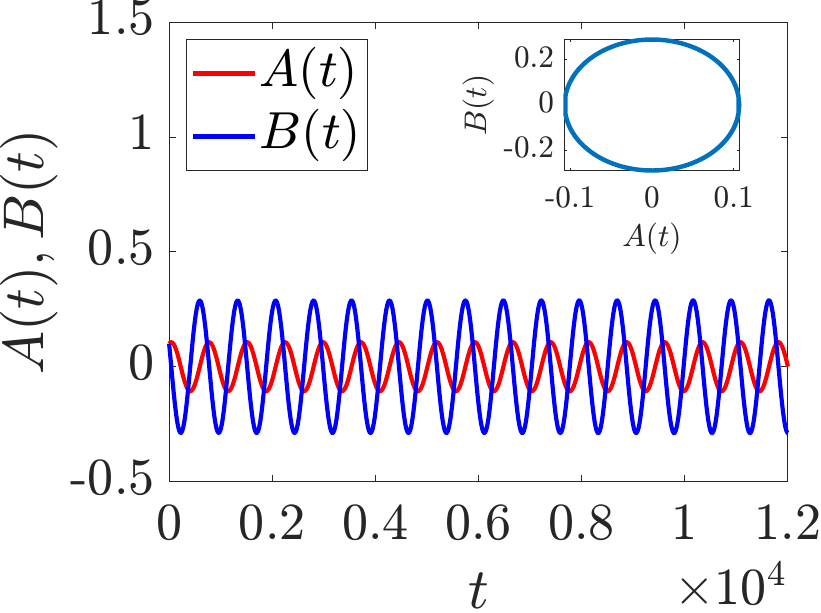}}

    \subcaptionbox{$V_0=7<V_{0c}=8.25,g=0$}{\includegraphics[scale=0.425]{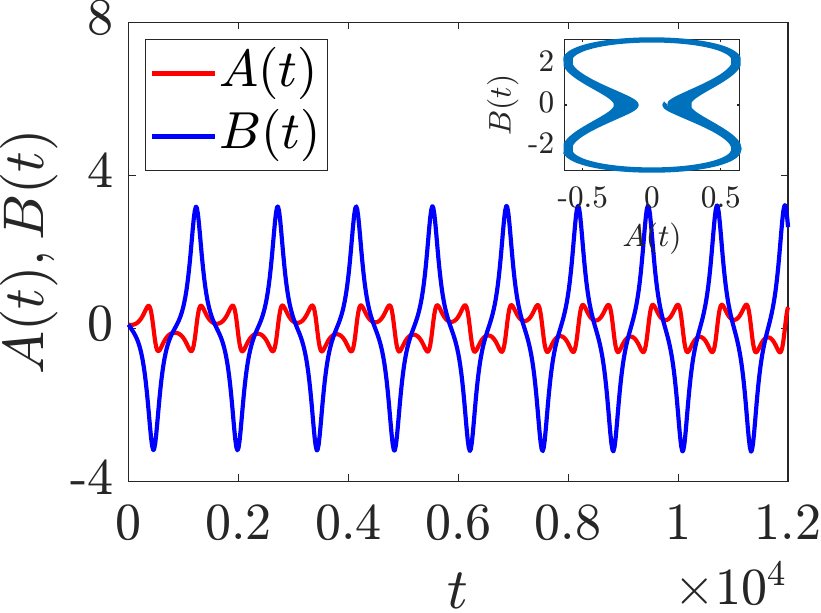}}
    \subcaptionbox{$V_0=9>V_{0c}=8.25,g=0$}{\includegraphics[scale=0.425]{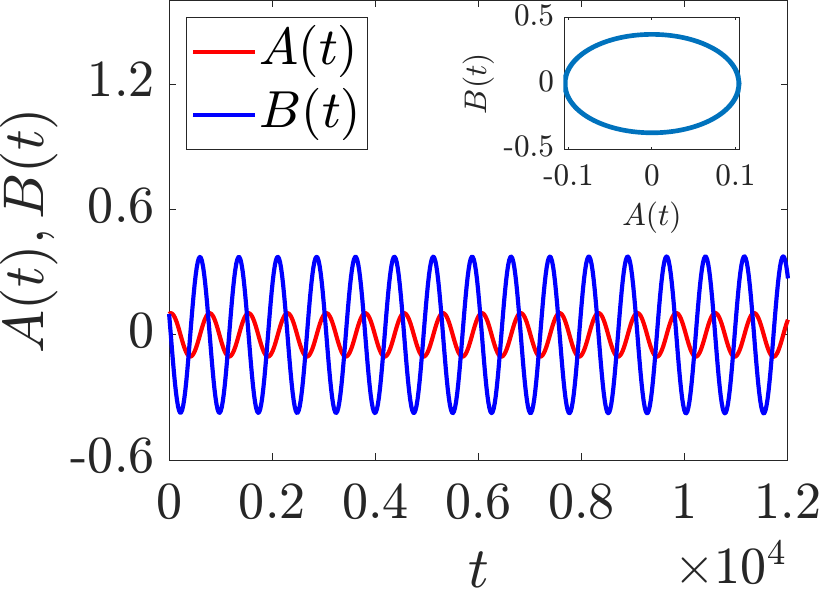}}
    \caption{Amplitude of $\langle\hat{X}\rangle$ for the simplified flow Equations with $V_{1}=V_{2}=0.$ For the first two panels (a) and (b) with $g=200$, $V_{0c}=5.95$. For panels (c) and (d) with $g=0$,$V_{0c}=8.25$. Below and above $V_{0c}$, a set of $V_0$ is appropriately selected to allow for the clear observation of a high amplitude to low amplitude transition, both with and without fast forcing. Other parameters are taken as: $\omega_p=1.0$, $\delta=0.1$, $\epsilon=0.11$, $\omega_0=0.5$ and $\Omega=5$ }
    \label{fig2}
\end{figure*}

However, as the nonlinearity increases, the threshold decreases, and the critical strength gradually decreases and asymptotically reaches zero as we turn on the external forcing; also, the parametric resonance zone ($V_0<V_{0c}$), where the erratic behavior of the trajectories and the fluctuation can be seen with high amplitude oscillation, shrinks. Figure~\ref{fig1}(b) shows a similar effect for three different fast forcing frequencies ($\Omega$) values.

\begin{figure*}[h]
    \subcaptionbox{$V_0=0.7,g=0$}{\includegraphics[scale=0.425]{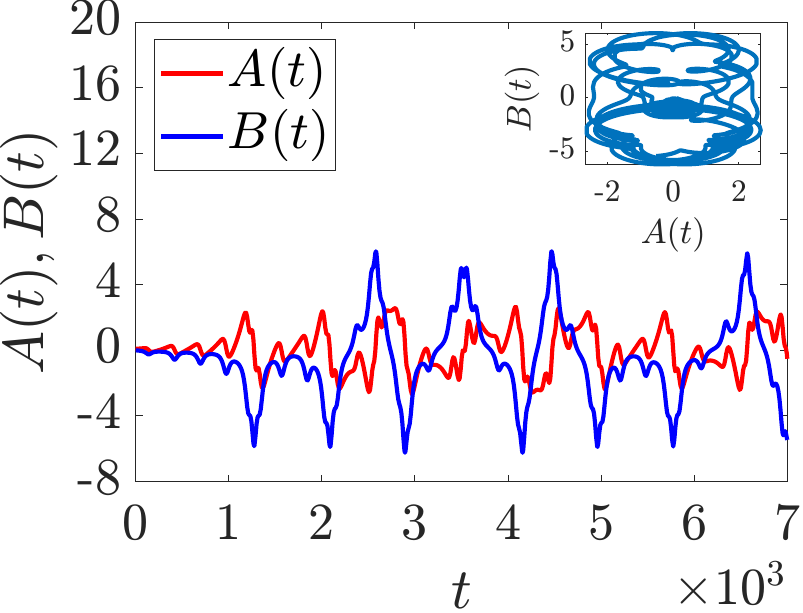}}
    \subcaptionbox{$V_0=0.7,g=0$}{\includegraphics[scale=0.425]{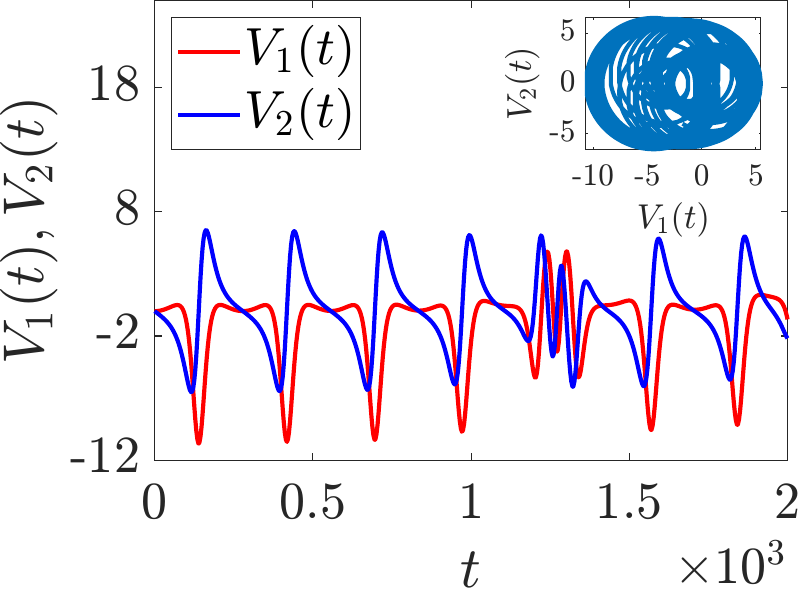}}
    \subcaptionbox{$V_0=0.7,g=0$}{\includegraphics[scale=0.425]{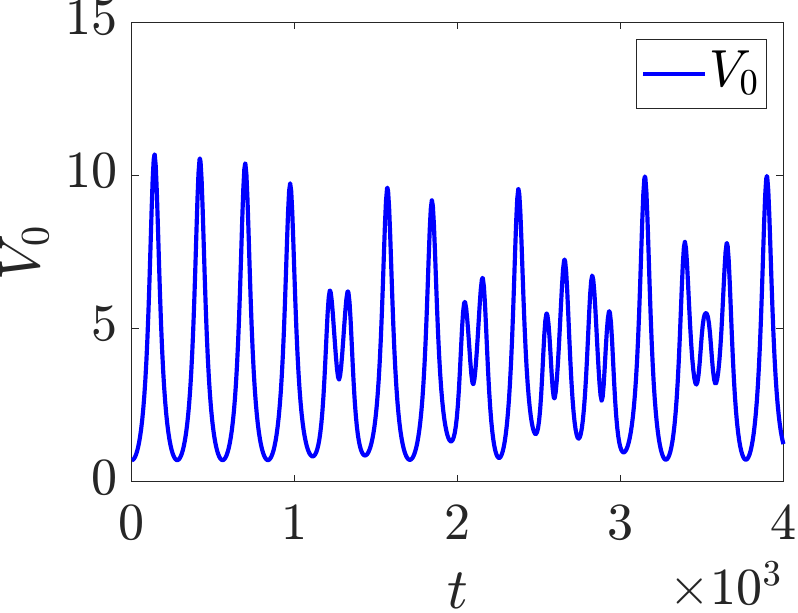}}
        
    \subcaptionbox{$V_0=0.7,g=200$}{\includegraphics[scale=0.425]{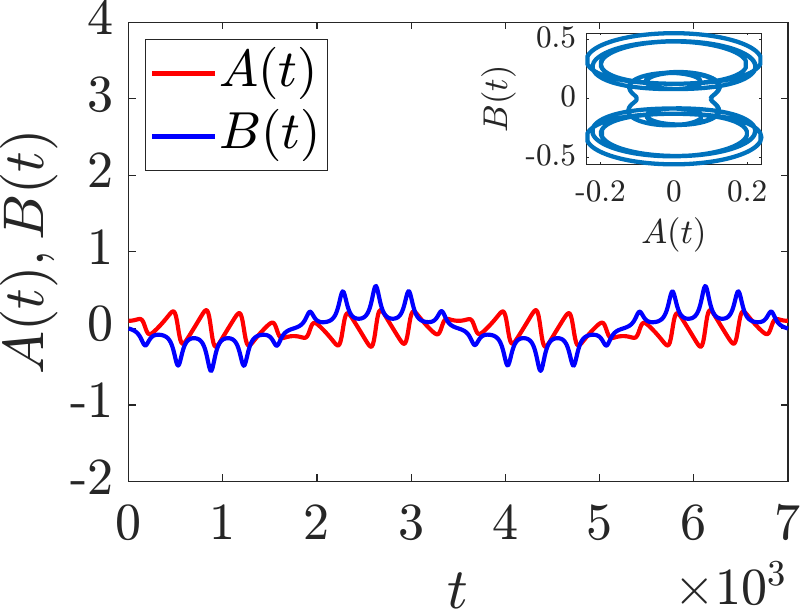}}
    \subcaptionbox{$V_0=0.7,g=200$}{\includegraphics[scale=0.425]{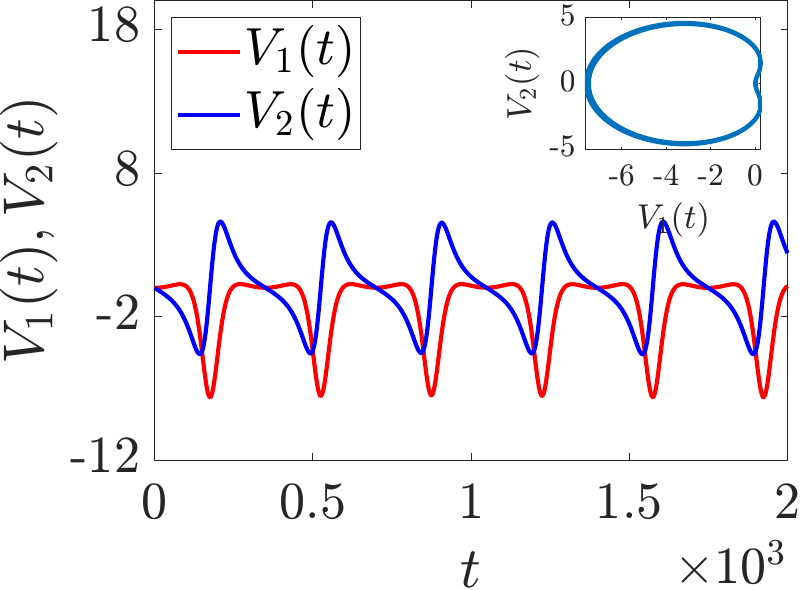}}
    \subcaptionbox{$V_0=0.7,g=200$}{\includegraphics[scale=0.425]{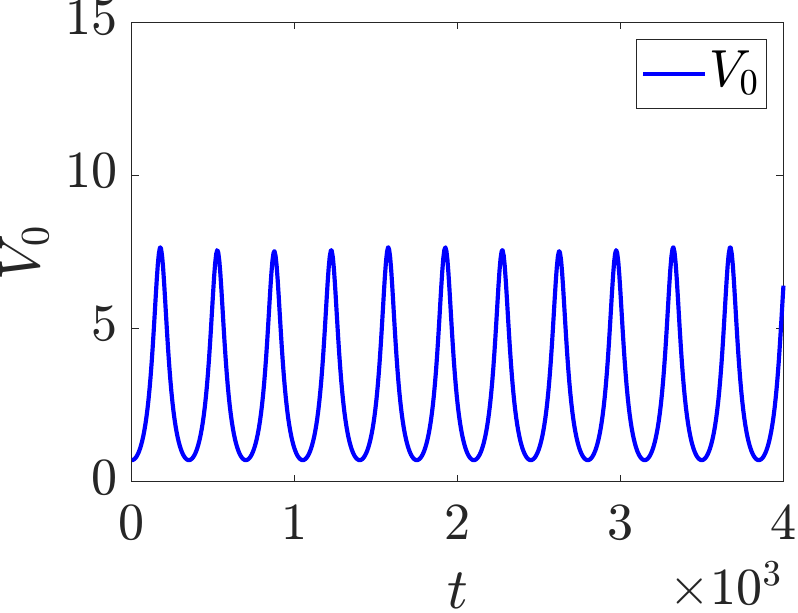}}
\caption{Dynamics of the flow equations Eqs.~\eqref{flow1}-\eqref{flow5} in the resonance zone (Zone A). The first row illustrates the high-amplitude, aperiodic oscillations of the variables \( (A, B) \), \( (V_1, V_2) \), and \( (V_0, t) \) in the absence of high-frequency signal (HFS), i.e., for \( g = 0 \). In contrast, the second row shows the suppressed and periodic oscillations of the same variables when the HFS is applied with \( g = 200 \). The initial conditions are \( A(0) = 0.2 \), \( B(0) = 0 \), \( V_0(0) = 0.7 \), \( V_1(0) = 0 \), and \( V_2(0) = 0 \). The other parameters are \( \lambda = 0.001 \), \( \epsilon = 0.11 \), \( \delta = 0.1 \), and \( \Omega = 5 \).}
\label{fig3}
\end{figure*}

For clarity, an expansion of the non-zero amplitude $B$ in Eq.\eqref{FPZoneA} in the limit : $\tilde{\lambda} \equiv \lambda/\omega_{0}^{2} \ll  2\Omega^{4}/ 3 g^{2}$ is demonstrated upto first order, 
\begin{equation}
 B^{2}= \frac{4\omega_{p}}{3\lambda}\Big[\epsilon ( \frac{\omega_{p}}{8} - \delta)\Big]  -4 V_{0} -\frac{2\omega_{p}g^{2}}{ \Omega^4}(\frac{\omega_{p}}{8} - \delta)\label{FPZoneAExp}
\end{equation}

The first term on the RHS is dominated by the classical $\lambda ^{-1}$ oscillations. The presence of quantum fluctuations in the second term reduces the magnitude of oscillations to a degree. But the quantum effect of external control via HFS is evident in the third term where remarkably the $\lambda^{-1}$ behavior is suppressed. The third term of the equation makes it evident that the greater the value of ($\Omega$), the greater the force required to observe the diminishing effect of the resonance zone around the fixed points. Accordingly, it is found that the interplay between nonlinearity-induced quantum fluctuations and the HFS can lead to a reduction of oscillation amplitude within the PSR. These qualitative arguments carry over to the fully general flow equations with all amplitudes being present,ie., Eqs.~\eqref{flow1}-\eqref{flow5}. They will be finally confirmed via direct numerical integration of the effective equations of motion for the average and variance as given by Eq.~\eqref{Moment2} in Sec.~\ref{sec5}.
In order to further understand how g affects $V_{0c}$ and, consequently, moment and fluctuation dynamics, the flow equations Eqs. (\ref{flow1} and \ref{flow2}) have been plotted in Fig.~\ref{fig2} with $V_1=V_2=0$ and $V_0$ treated as a parameter only. When $g=0$ and $g=200$, the values of $V_{0c}=8.25$ and $5.9$5 are obtained from equation Eq.~\eqref{Vcritical} respectively. The resonance zone is confirmed by the value of $\delta=-0.1$ for $\omega_p=1.0$. In this way, one can anticipate an oscillation with a high amplitude when $V_0<V_{0c}$ and a low amplitude when $V_0>V_{0c}$.  Additionally, it is observed that by turning on g, the critical value of quantum fluctuation can be reduced, increasing the area of regular low-amplitude oscillation.
For further clarification,  in order to validate how the HFS qualitatively predicts the diminishing effect of the amplitude of the moment and the fluctuation itself, as well as to support the quantitative description provided in Fig.~\ref{fig2}, the simulation on the five flow equations Eqs.~\eqref{flow1}-\eqref{flow5} have been carried out. Figure~\ref{fig3} shows how crucial initial conditions are in regulating the dynamics. The initial value of $V_0$ in the flow equation Eq.~\eqref{flow3} closely resembles the role of criticality as in Eq.~\eqref{Vcritical}, where $V_0$ only acts as a parameter. We don't expect an analytic closed form formula like in Eq.~\eqref{Vcritical}, as now $V_0, V_1, V_2$ all are slow functions of time and participate in the moment dynamics over time. 
Two rows of the dynamics of $(A,B), (V_1,V_2)$ and $(V_0,t)$ are illustrated in Fig.\ref{fig3}. The dynamics in the absence of HFS $(g=0)$, where the erratic and irregular large amplitude dynamics predominate, are displayed in the first row ($a,b,c$). However, the dynamics favourably settle down to lower values in the second row ($d,e,f$), when the HFS is driving the system $(g=200)$, and a low amplitude periodic behavior is more prominent. Intuitively, Eq.~\eqref{FPZoneAExp} also clearly indicates that the amplitude of oscillations may be reduced by a rather significant factor by increasing $g$ in the presence of $V_ 0$.

\begin{figure*}[h]
    \subcaptionbox{$g=0$}{\includegraphics[scale=0.425]{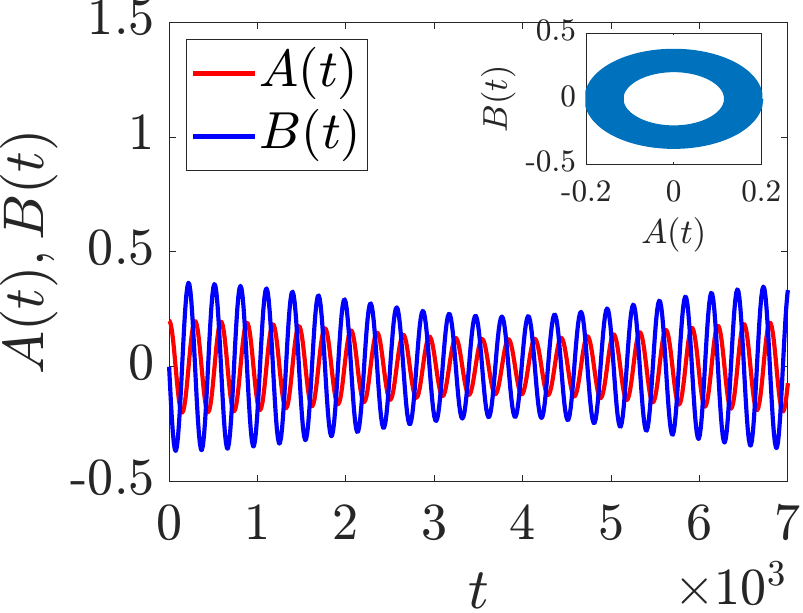}}
    \subcaptionbox{$g=0$}{\includegraphics[scale=0.425]{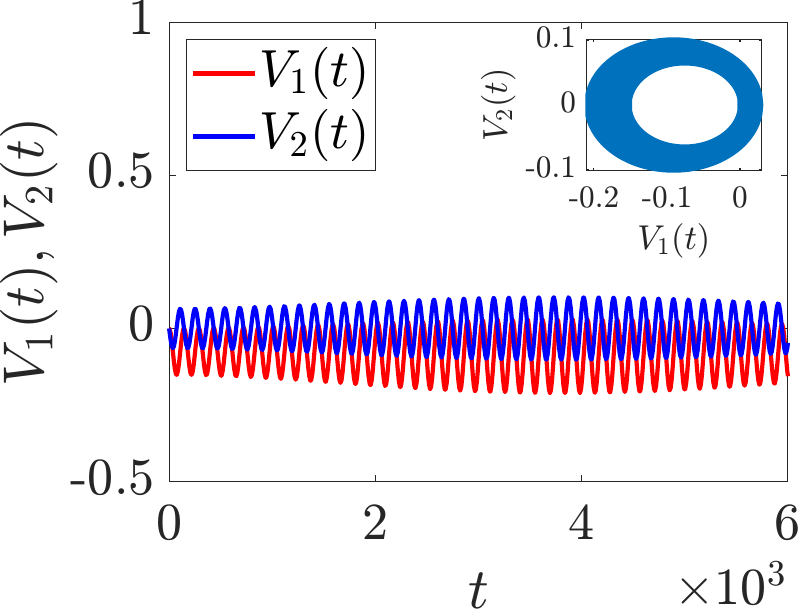}}
    \subcaptionbox{$g=0$}{\includegraphics[scale=0.425]{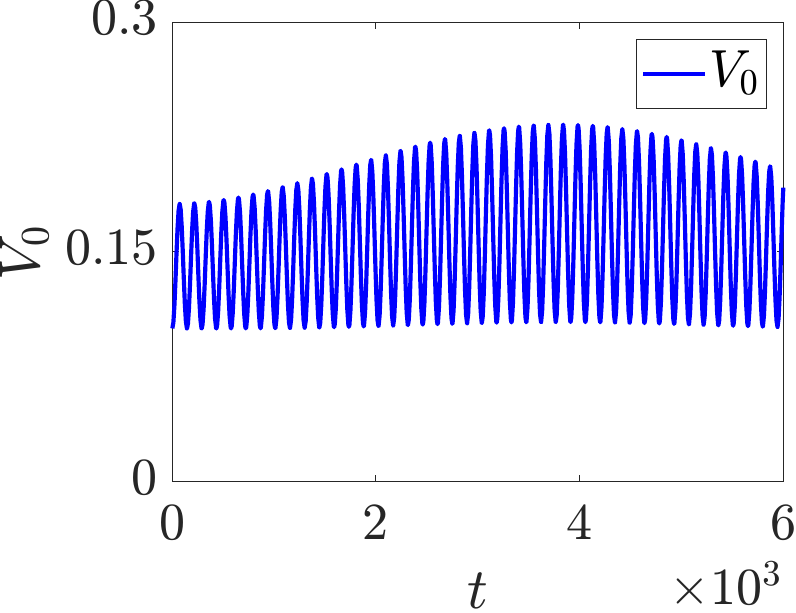}}
        
    \subcaptionbox{$g=400$}{\includegraphics[scale=0.425]{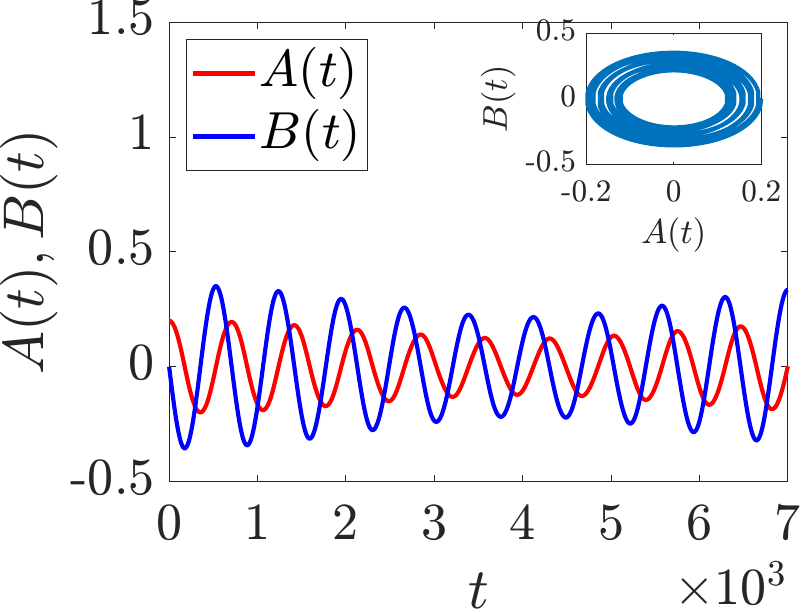}}
    \subcaptionbox{$g=400$}{\includegraphics[scale=0.425]{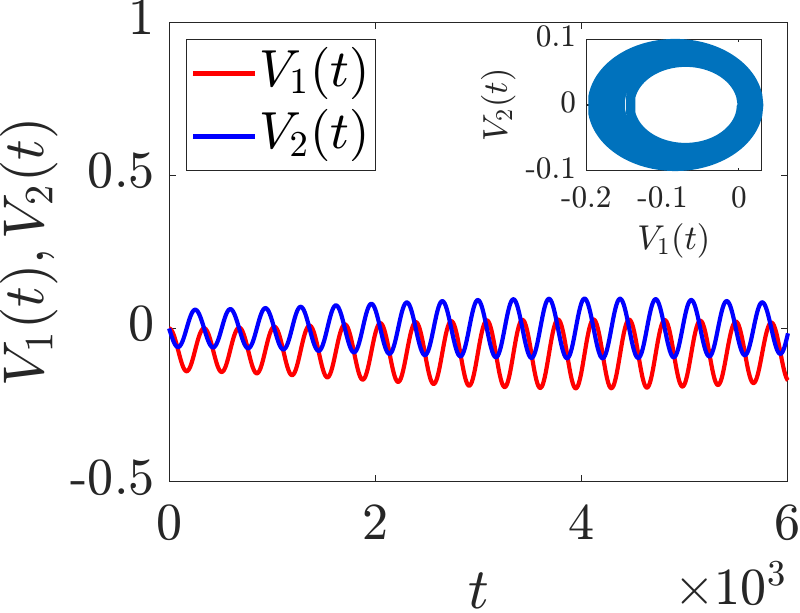}}
    \subcaptionbox{$g=400$}{\includegraphics[scale=0.425]{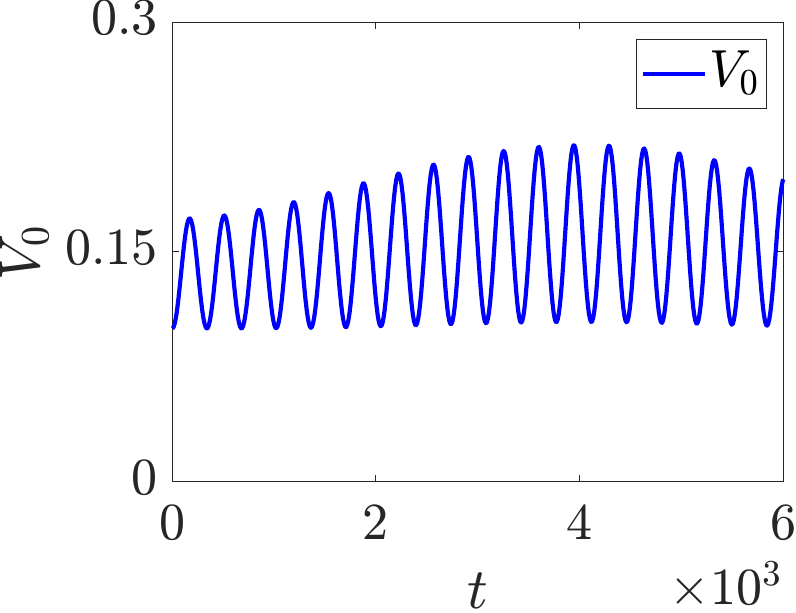}}
    \caption{Dynamical evolution in Zone $B_2$ for \( \delta = 0.225 > \omega_p/8 \), illustrating the effect of fast forcing using Eqs.\eqref{flow1}-\eqref{flow5}. The first row depicts the dynamics of \( (A, B); (V_1, V_2) \) and \( (V_0,t) \) in the absence of fast forcing (\( g = 0 \)), showing low-amplitude oscillations. The second row displays the same variables under strong fast forcing (\( g = 400 \)), where the oscillation period elongates for each variable due to the reduction in the effective parametric strength \( \epsilon_r \).}
    \label{fig:zoneB_posdelta}
\end{figure*}

\begin{figure*}[h]
\subcaptionbox{$g=0$}{\includegraphics[scale=0.425]{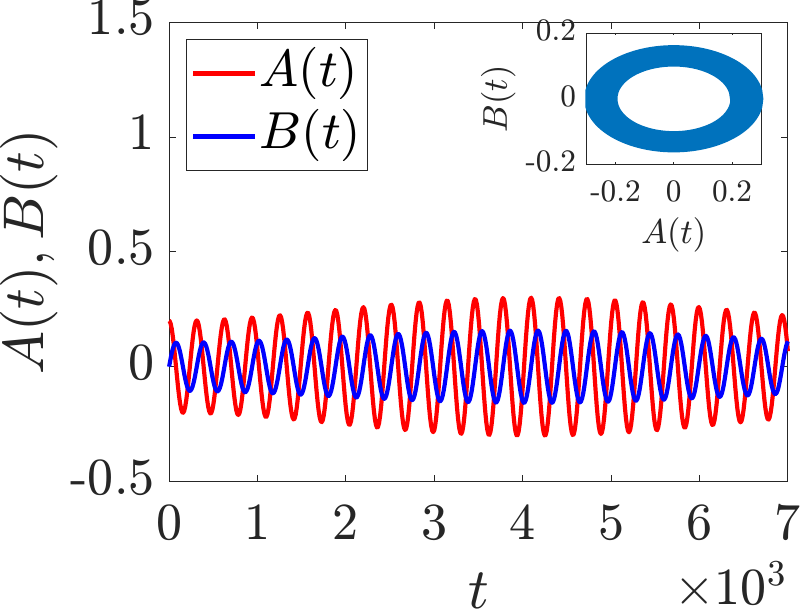}}
\subcaptionbox{$g=0$}{\includegraphics[scale=0.425]{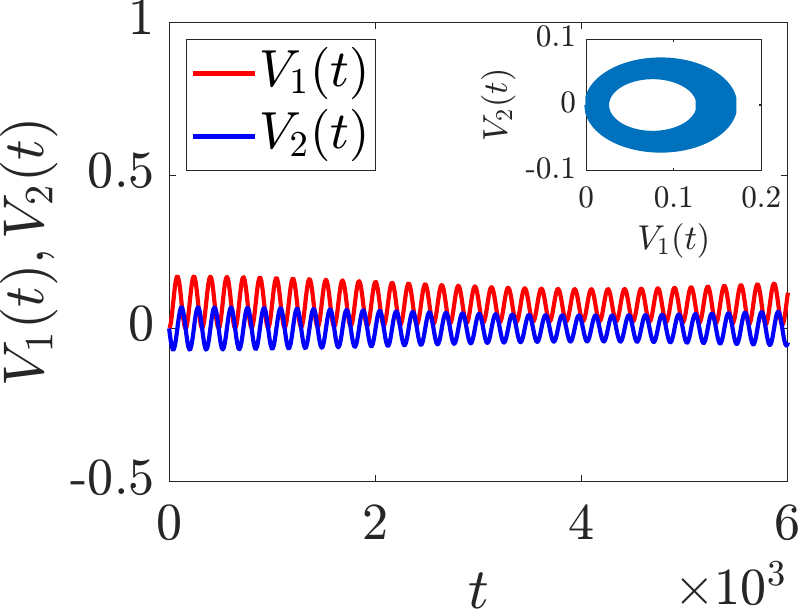}}
\subcaptionbox{$g=0$}{\includegraphics[scale=0.425]{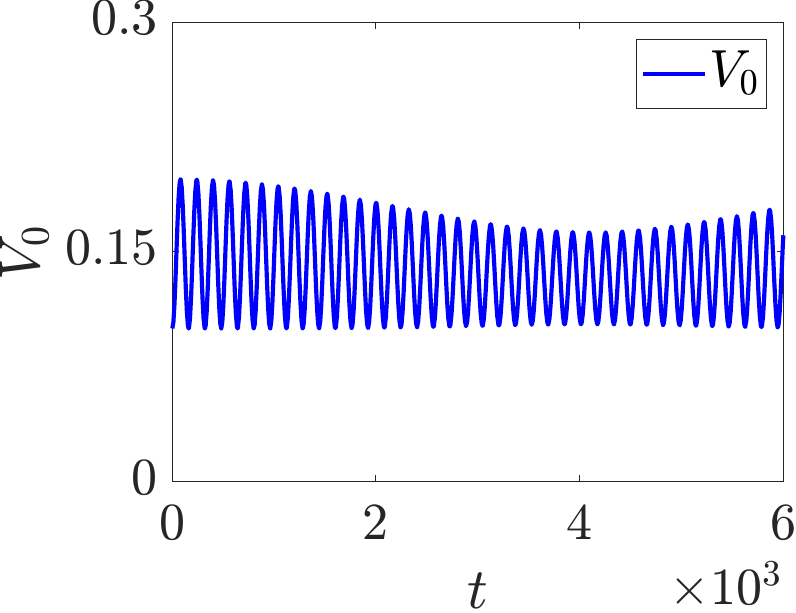}}
    
\subcaptionbox{$g=400$}{\includegraphics[scale=0.425]{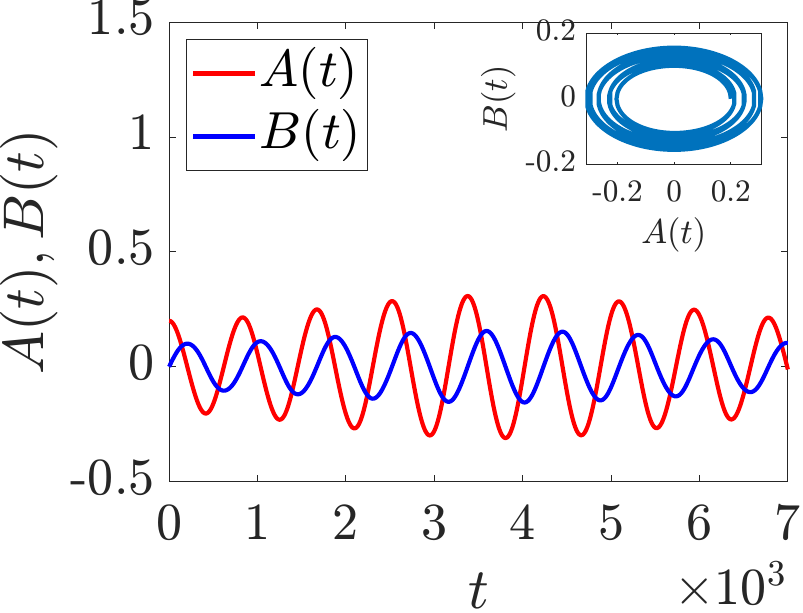}}
\subcaptionbox{$g=400$}{\includegraphics[scale=0.425]{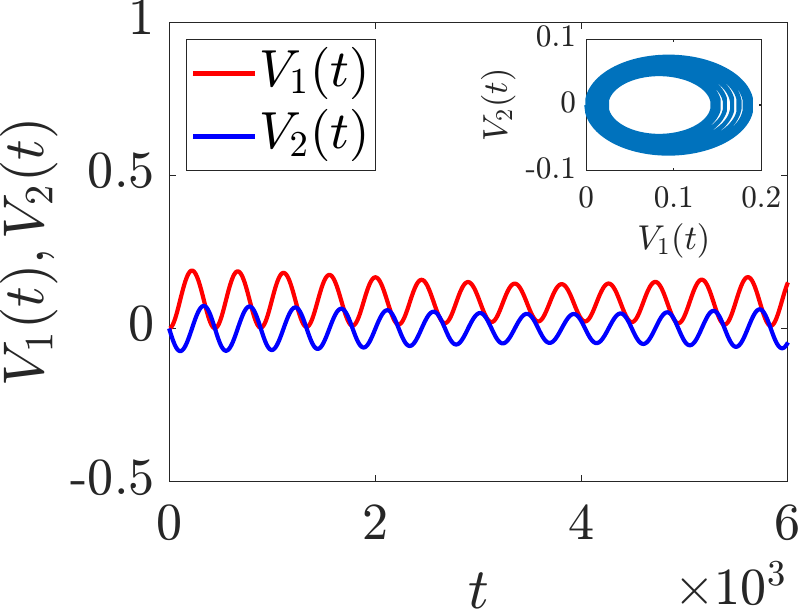}}
\subcaptionbox{$g=400$}{\includegraphics[scale=0.425]{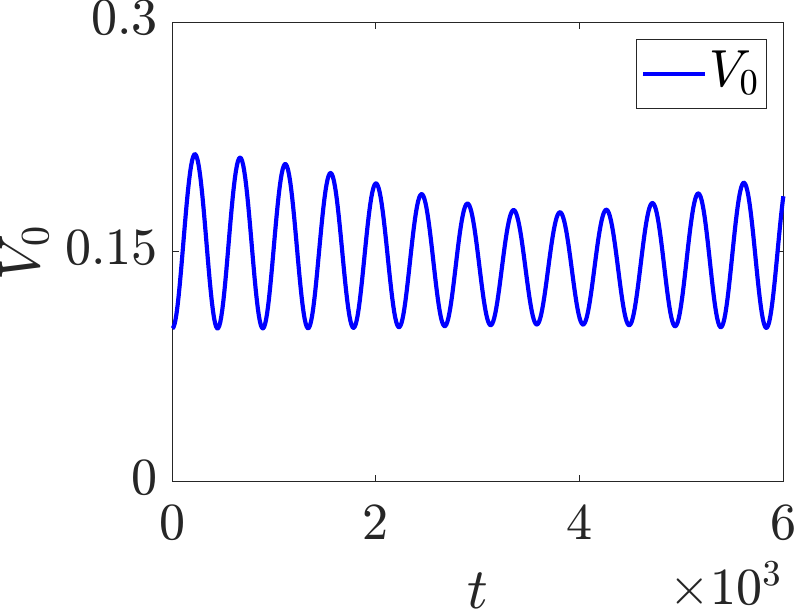}}
\caption{Dynamical evolution in Zone $B_1$ for \( \delta = -0.225 < -\omega_p/8 \), illustrating the effect of fast forcing by using Eqs.~\eqref{flow1}-\eqref{flow5}. The first row depicts the dynamics of \( (A, B); (V_1, V_2) \) and \( (V_0,t) \) in the absence of fast forcing (\( g = 0 \)), showing low-amplitude oscillations. The second row displays the same variables under strong fast forcing (\(g=400 \)). The presence of HFS leads to an extended oscillation period, highlighting the modulation of dynamics via the HFS.}

\label{fig:zoneB_negdelta}
\end{figure*}

\emph{Zone B}: The presence of HFS away from the resonance zone also has interesting implications on the effect of quantum fluctuations in our model. Again, to gain a level of analytical insight, the flow equations can be investigated first in sub \textit{Zone $B_1$}, keeping in mind that in this zone $\delta > \omega_{p}/8 $, a trivial fixed point, $A=0, B=0$ exists classically and small periodic oscillations dominate around a centre. To include quantum mechanical fluctuations, we can constrain $V_{2}=0$ and we find that the fixed point transforms into a fixed curve given by $A=0, B=0, V_{2}=0$ but, $V_{1} + \frac{\omega_{p}}{8 \delta}V_{0} + \frac{9 \lambda}{2 \epsilon_{r} \delta \omega_{p}}V_{1}V_{0} =0$ \cite{sarkar2020nonlinear}. It is to be noted also that parameter values can be chosen (and relevant to the numerical solutions in our discussion later) such that the coefficient of the third term is markedly smaller than 1 and the curve can be approximated by a straight line. Thus, for $g=0$, as reported in Ref.~\cite{sarkar2020nonlinear},  as well as for non-zero values of $g$, the effect of $V(t)$ is to smear the classical trivial fixed point into a curve (or a straight line).
Notice, via the presence of $\epsilon_{r}\propto 1/(\omega_{0}^{2} + \frac{3\lambda g^{2}}{2 \Omega^4})$, the nature of the fixed curve can be modulated since the effective parametric strength is dependent on the HFS strength $g$. In fact, a closer look at how the curve behaves for different values of $g$ tells us the following: we can make sure that a wider range of the curve fits the constraint $V(t)>0$ (since, it is the variance of an observable) and yet retains the ability of producing stable small oscillations away from the resonance zone.
The flow equations (\ie, Eqs.~\eqref{flow1} and~\eqref{flow2}) offer additional insight into the role of the parametric strength $\epsilon_r$ in shaping the dynamics of $A(t')$, where $t'=\epsilon_r t$, is the slow time. In the presence of small nonlinearity, the evolution of $A(t')$ can be approximated as proportional to $\cos\left(\epsilon_r\sqrt{\delta^2 - \left(\omega_p/8\right)^2} t + \phi\right)$, with a corresponding time period given by $T \approx 2\pi/\epsilon_r\sqrt{\delta^2-(\omega_p/8)^2}$. Importantly, since the HFS strength $g$ contributes to the parametric strength $\epsilon_r$, an increase in $g$ leads to a reduction in $\epsilon_r$, thereby elongating the time period of the modulated amplitude.

Figure~\ref{fig:zoneB_posdelta} illustrates the dynamics in Zone $B_2$, corresponding to the parameter region where \( \delta > \omega_p/8 \). Here, we choose \( \delta = 0.225 \) and fix all other parameters as \( \lambda = 0.001 \), \( \Omega = 5 \), \( \omega_0 = 0.5 \), \( \omega_p = 1.0 \), and \( \epsilon = 0.11 \). The initial conditions are taken as \( V_0 = 0.1 \), \( V_1 = 0 \), and \( V_2 = 0 \). In the first row of Fig.~\ref{fig:zoneB_posdelta}, we depict the evolution of the system in the absence of HFS (i.e., \( g = 0 \)), where a low-amplitude oscillation is observed with a time period of \( T \approx 305.3 \), as estimated from the flow equation approximation. When the HFS is switched on with strength \( g = 400 \) (shown in the second row), although the qualitative nature of the oscillations remains largely unaltered, the time period increases significantly to \( T \approx 774.3 \). This shift results from a decrease in the effective parametric strength \( \epsilon_r \), whose expression is given by
\[
\epsilon_r = \frac{\epsilon \omega_0^2}{\omega_0^2 + (3\lambda g^2/2 \Omega^4
)}.
\]
Substituting the numerical values gives \( \epsilon_r \approx 0.0434 \) for \( g = 400 \), compared to \( \epsilon_r = 0.11 \) for \( g = 0 \). This clearly demonstrates how increasing the HFS strength effectively stretches the time scale of the modulated amplitude, outside the main resonance zone.\\

In Figure~\ref{fig:zoneB_negdelta}, which illustrates the dynamics in the other Zone $B_1$ corresponding to \( \delta < -\omega_p/8 \), we observe a similar phenomenon as described in Fig.~\ref{fig:zoneB_posdelta}. Here, the detuning parameter is chosen as \( \delta = -0.225 \), with all other system parameters and initial conditions kept the same as before. The top row of Fig.~\ref{fig:zoneB_negdelta} shows the system's evolution without HFS (\( g = 0 \)), where a low-amplitude oscillatory state emerges with a time period again around \( T \approx 305.3 \). Upon introducing the fast forcing (\( g = 400 \)) in the bottom row, a marked elongation in the time scale of oscillation is observed, with the period increasing to approximately \( T \approx 774.3 \), consistent with the reduction in effective parametric strength \( \epsilon_r \approx 0.0434 \). This demonstrates the general applicability of the time-scale separation technique in modulating quantum fluctuation dynamics and validates that the effect of HFS on modulation frequency is robust across both positive and negative detuning regimes outside the resonance zone.

To summarize, it is found that quantum fluctuations and the requirement of positive variance $V(t)$ define a narrow window of stable fixed points outside the resonance zone. This window can be widened or shrinked through the application of high-frequency signals (HFS). Additionally, increasing the forcing strength is observed to stretch the time period of the modulation amplitude in both dynamical zones. Finally, the findings derived from the flow equations are compared through comparison with results obtained from the effective equations, which are numerically solved and discussed in the following section.

\begin{figure*}[h]
    \subcaptionbox{$g=0$}{\includegraphics[scale=0.425]{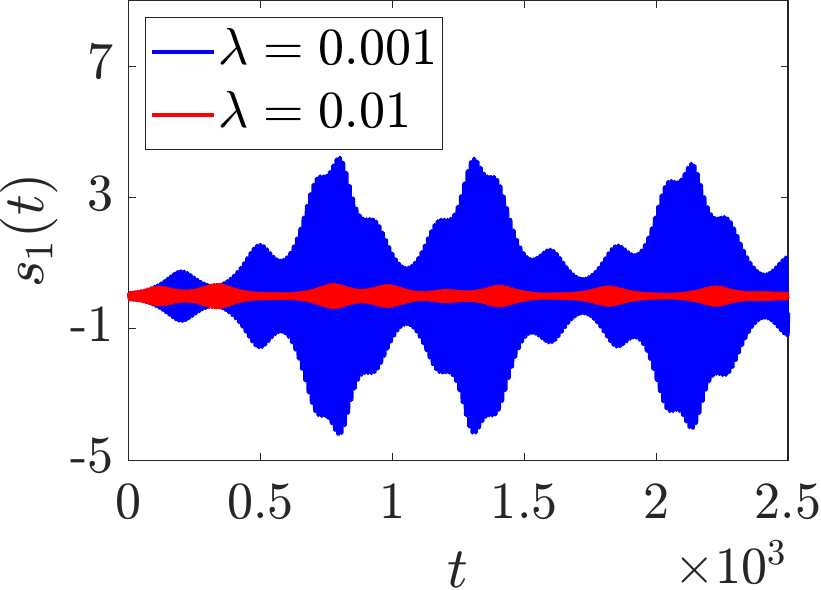}}
    \subcaptionbox{$g=60$}{\includegraphics[scale=0.425]{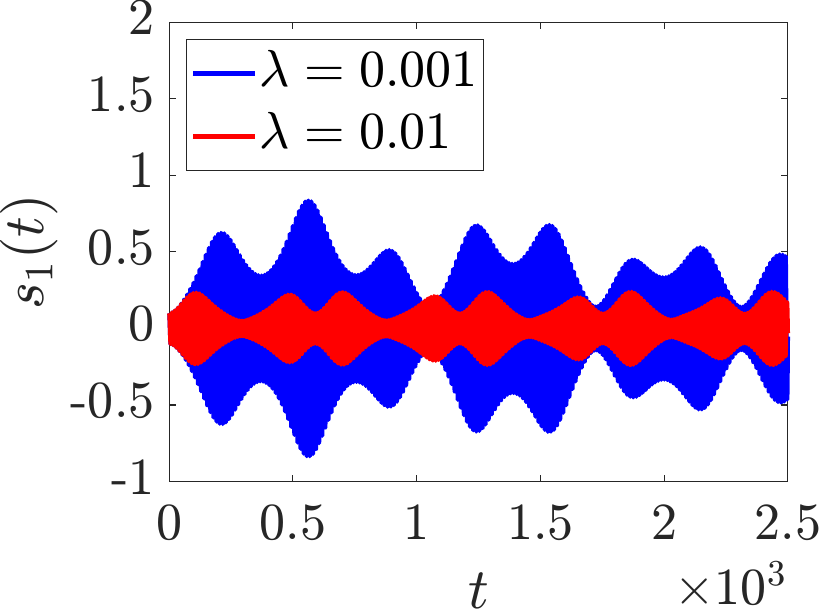}}
    \subcaptionbox{$g=100$}{\includegraphics[scale=0.425]{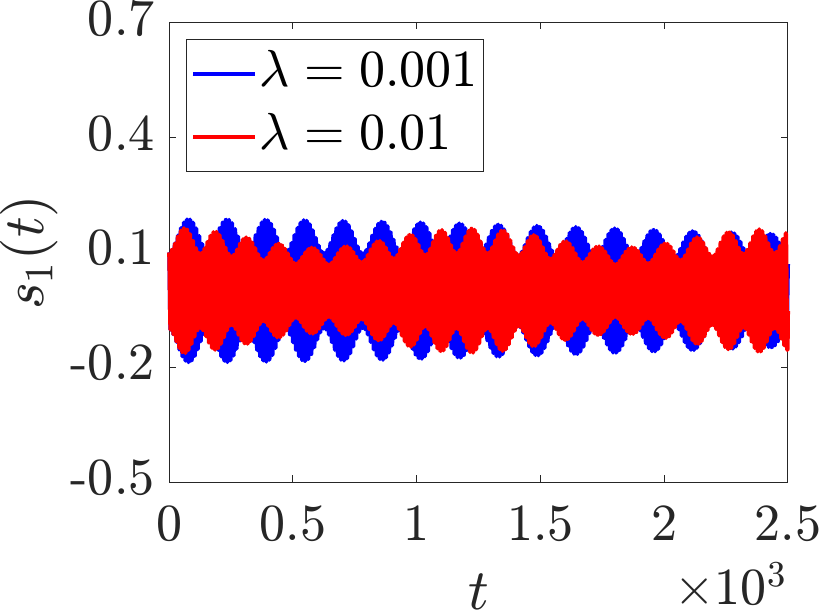}}
        
    \subcaptionbox{$g=0$}{\includegraphics[scale=0.425]{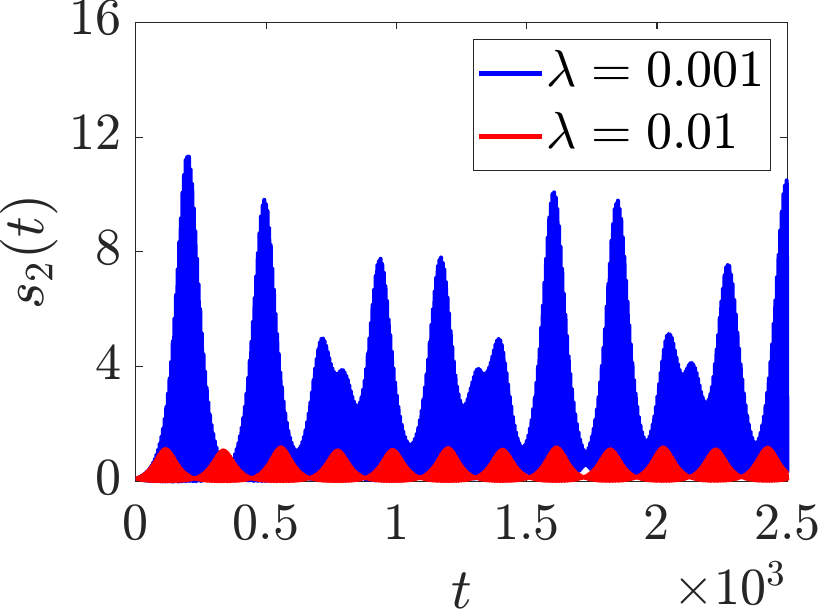}}
    \subcaptionbox{$g=60$}{\includegraphics[scale=0.425]{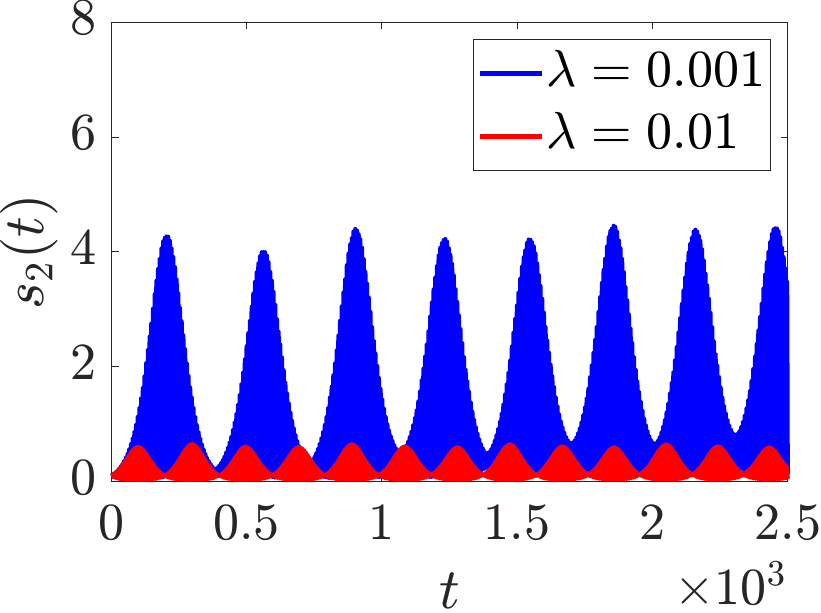}}
    \subcaptionbox{$g=100$}{\includegraphics[scale=0.425]{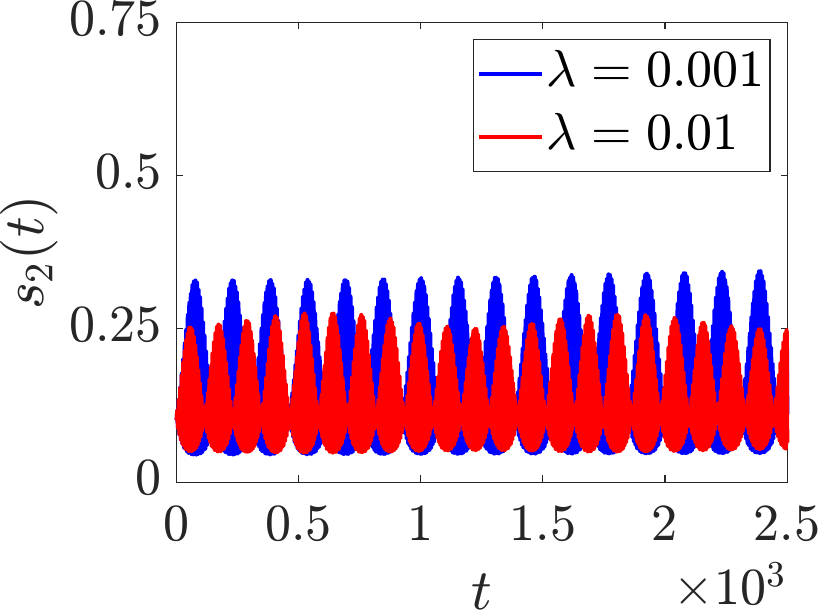}}

 \caption{The original (effective) dynamics (Eq.~\eqref{Moment2}) in the resonance zone (\( \omega_0 = \omega_p/2 = 0.5 \)) are presented. The first row shows the evolution of the effective average \( \langle X(t) \rangle \) (or \( s_1 \)) for varying HFS strengths \( g = 0, 60, 100 \) and two different values of \( \lambda \). The second row illustrates the corresponding effective dynamics of \( V(t) \) (or \( s_2 \)) under the same conditions. The other system parameters are \( \epsilon = 0.11 \), \( \Omega = 5.0 \).The initial conditions used are \( s_1(0) = 0.1,\, \dot{s}_1(0) = 0 \) and \( s_2(0) = 0.1,\, \dot{s}_2(0) = 0,\, \ddot{s}_2(0) = 0 \).}
 \label{fig:original_z1}
\end{figure*}

\begin{figure*}
     \subcaptionbox{$\lambda=0.001$}{\includegraphics[scale=0.425]{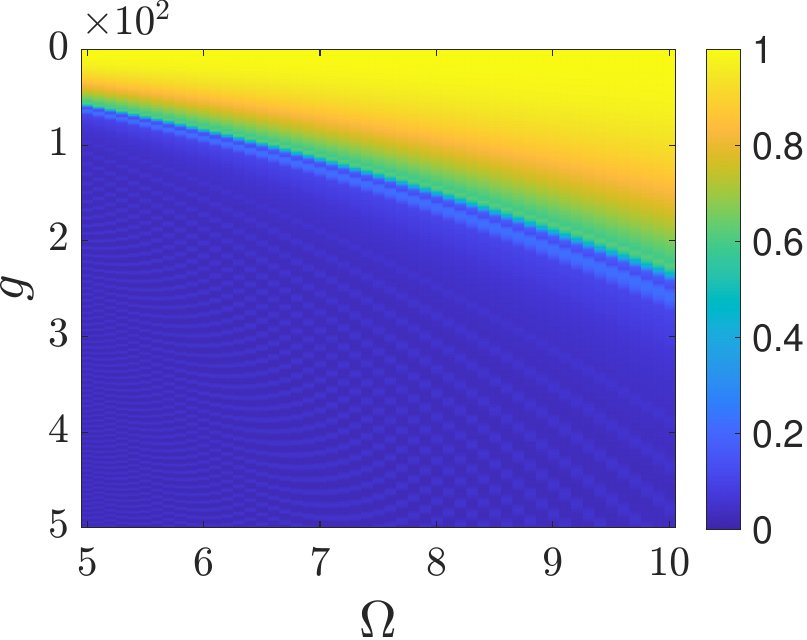}}
    \subcaptionbox{$\lambda=0.010$}{\includegraphics[scale=0.425]{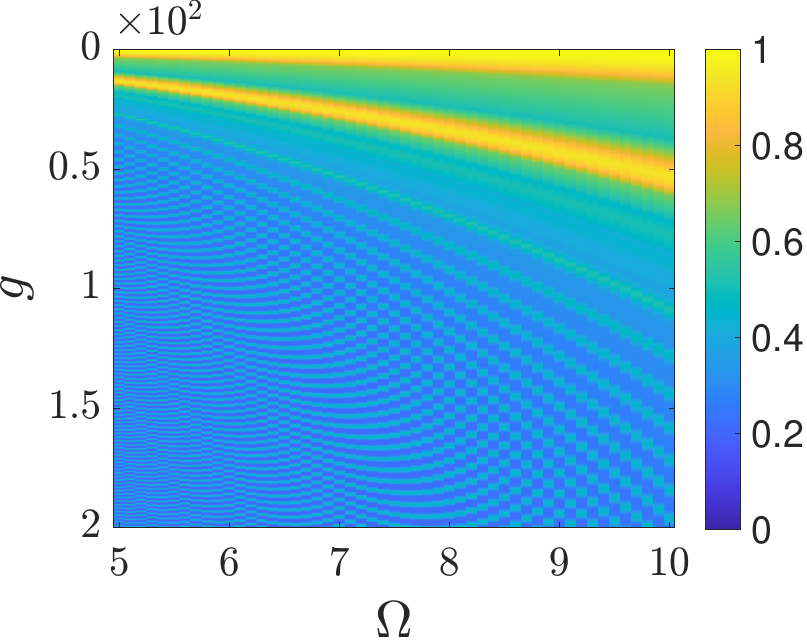}}
    \subcaptionbox{$\lambda=0.100$}{\includegraphics[scale=0.425]{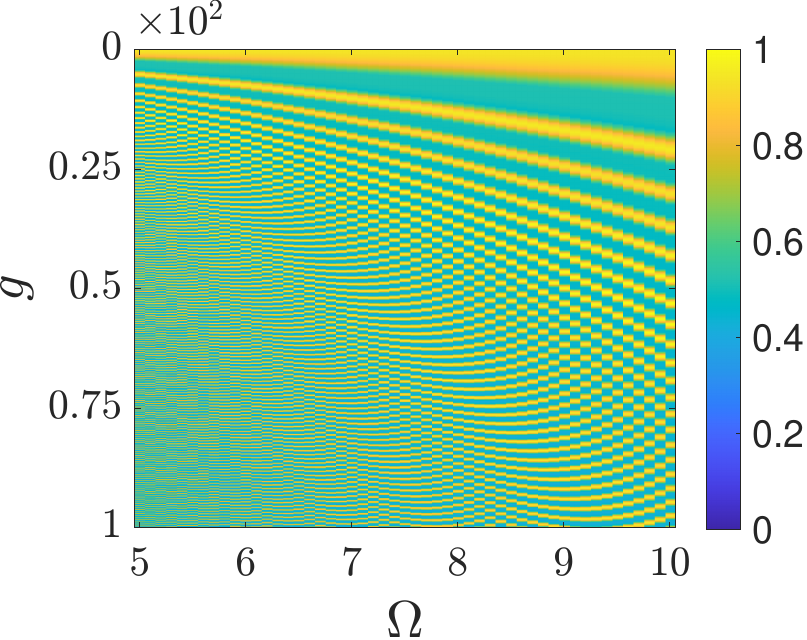}}
    \caption{Parametric region plot of the maximum amplitude of the effective quantum average \( \langle X \rangle \) (denoted as \( s_1 \)) in the \( g \)-\( \Omega \) plane for three different values of the nonlinearity parameter \( \lambda \) by Eq.~\eqref{Moment2}. The figure demonstrates that for weak nonlinearity, increasing the HFS strength \( g \) effectively suppresses the amplitude of \( s_1 \). However, in the strong nonlinearity regime, this suppression becomes less effective. The color bar indicates the maximum value of \( \langle X \rangle \). The parameters are set to \( \omega_p = 1 \), \( \omega_0 = 0.4 \), and \( \epsilon = 0.11 \). The initial conditions used are \( s_1(0) = 0.1,\, \dot{s}_1(0) = 0 \) and \( s_2(0) = 0.1,\, \dot{s}_2(0) = 0,\, \ddot{s}_2(0) = 0 \).}

    \label{fig:parametric_order}
\end{figure*}

\begin{figure*}[h]
    \subcaptionbox{$g=0$}{\includegraphics[scale=0.425]{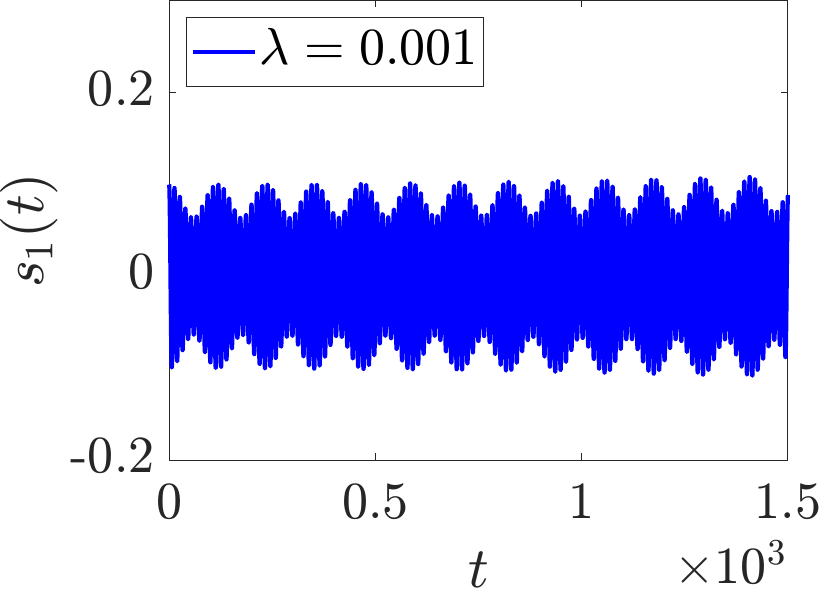}}
    \subcaptionbox{$g=74$}{\includegraphics[scale=0.425]{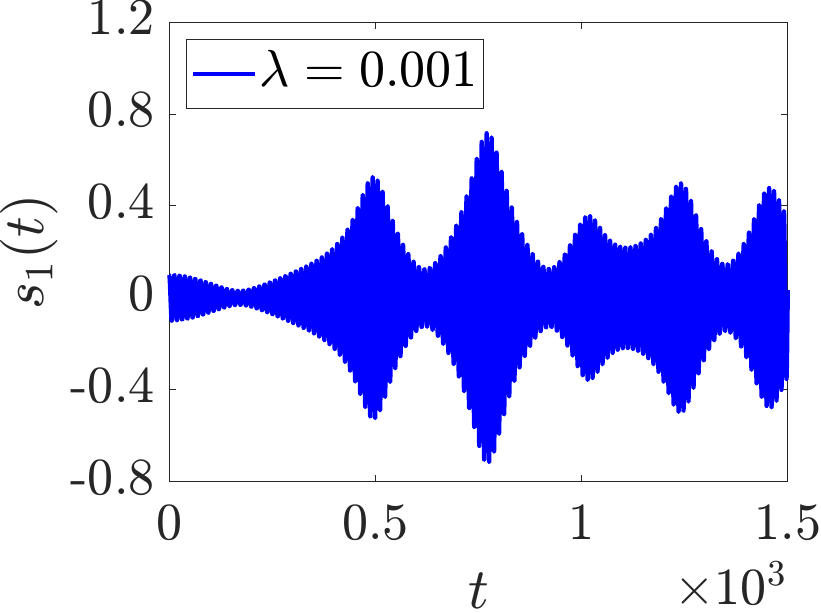}}
    \subcaptionbox{$g=134$}{\includegraphics[scale=0.425]{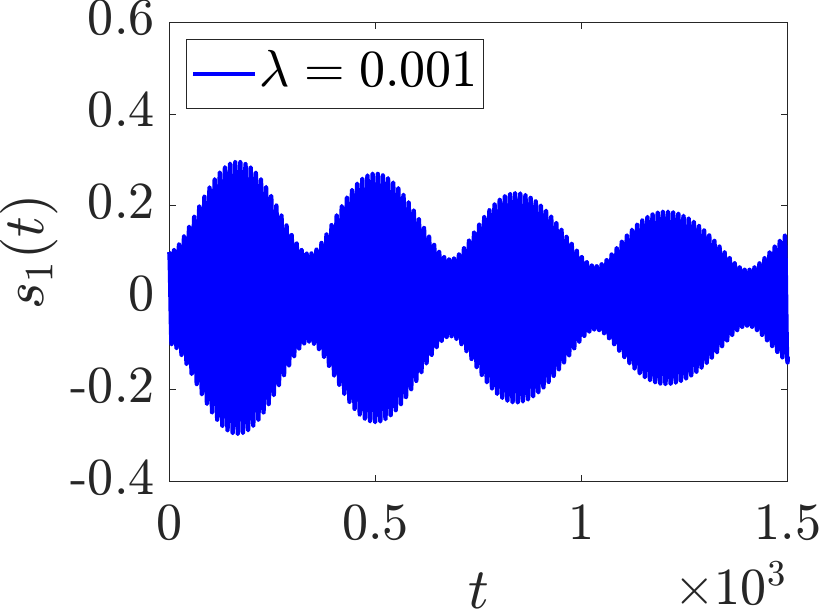}}
        
    \subcaptionbox{$g=0$}{\includegraphics[scale=0.425]{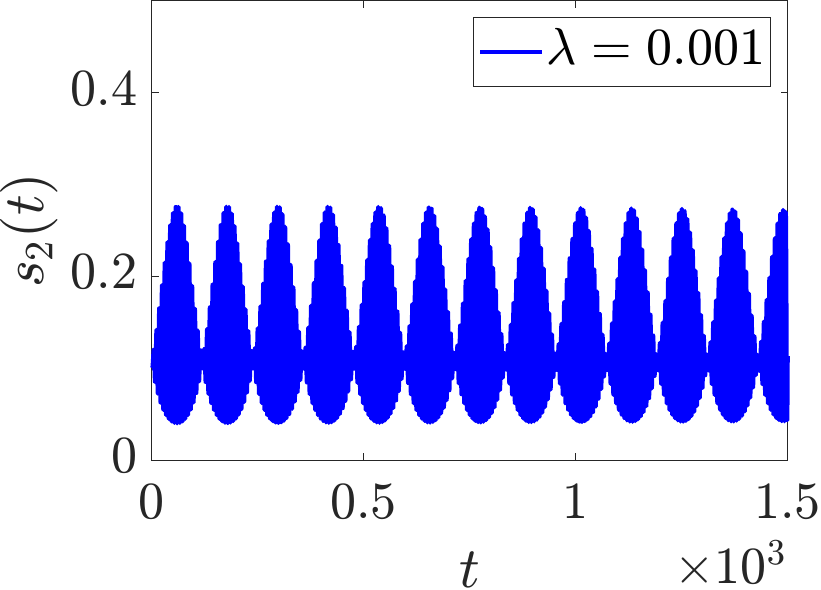}}
    \subcaptionbox{$g=74$}{\includegraphics[scale=0.425]{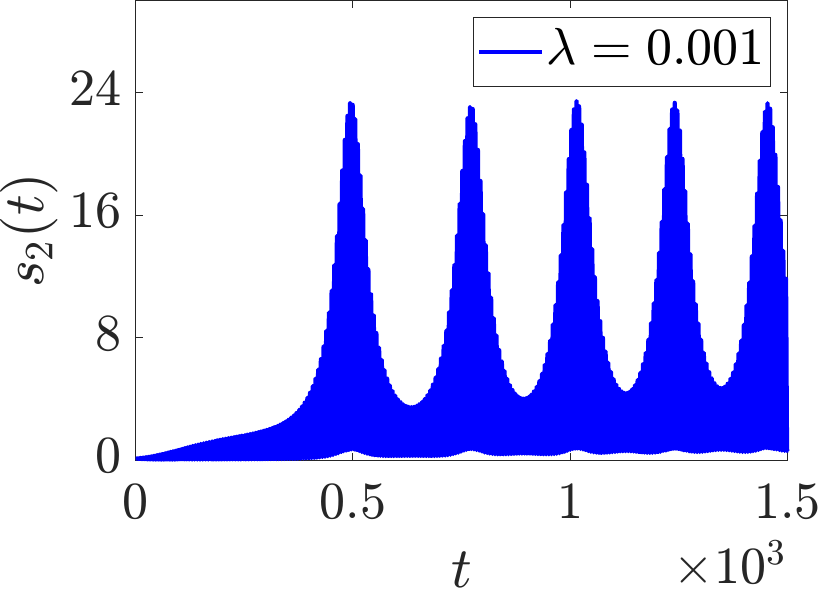}}
    \subcaptionbox{$g=134$}{\includegraphics[scale=0.425]{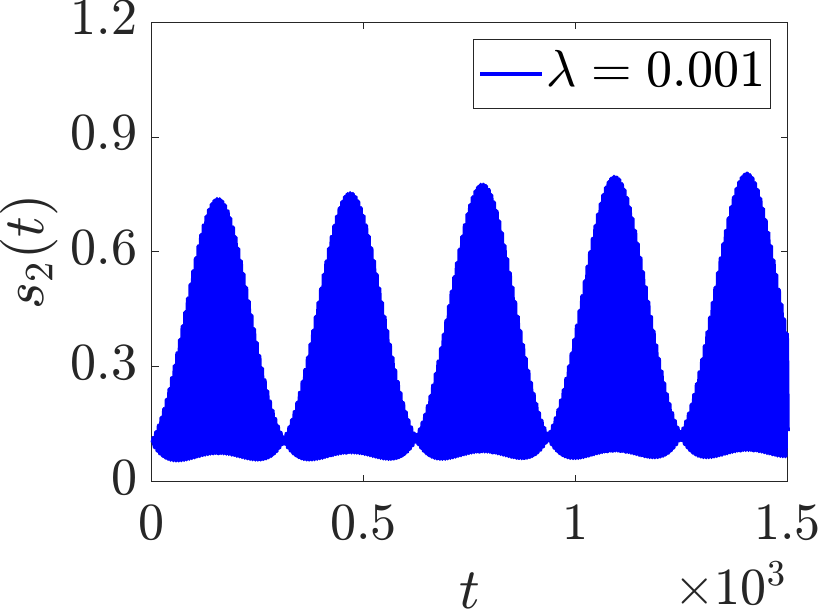}}
\caption{The original (effective) dynamics (\ie, Eq.~\eqref{Moment2}) in the \emph{Zone $B_1$} (\( \omega_0 < \omega_p/2 \)) are presented. The first row shows the evolution of the effective average \( \langle X(t) \rangle \) (or \( s_1 \)) for varying HFS strengths \( g = 0, 74, 134 \). The second row illustrates the corresponding effective dynamics of \( V(t) \) (or \( s_2 \)) under the same conditions. The other system parameters are \( \epsilon = 0.11 \), \( \omega_0 = 0.47 \),\( \omega_p = 1.0 \) and \( \Omega = 5.0 \). Dynamics shows transitions from resonant to low-amplitude periodic oscillations as the external strength $g$ is varied.
 The initial conditions used are \( s_1(0) = 0.1,\, \dot{s}_1(0) = 0 \) and \( s_2(0) = 0.1,\, \dot{s}_2(0) = 0,\, \ddot{s}_2(0) = 0 \).}
\label{fig:zoneB1}
\end{figure*}

\newpage
\section{Numerical results and discussions}
\label{sec5}
The numerical results for \emph{Zone A}, corresponding to the resonance region, are discussed first. Subsequently, the results of \emph{Zone $B_1$} where $\delta > \omega_p/8$ and \emph{Zone $B_2$} where $\delta < -\omega_p/8$, both of which lie outside the resonance zone, are also addressed respectively.

The numerical integration of the effective dynamics Eq.~\eqref{Moment2} for a weak ($\lambda=0.001$) and a relatively strong ($\lambda=0.01$) nonlinearity is shown in Fig.~\ref{fig:original_z1}. In contrast, when the nonlinearity increases, a small fluctuation can easily cross the threshold of criticality and regularize the dynamics. In the absence of HFS, for weak nonlinearity, critical fluctuation is high, and large amplitude aperiodic oscillation appears to occur in the system. Panels ($a,d$) show this. It is evident that the dynamics of \( s_1 \) (or \( \langle X \rangle \)) in Fig.~\ref{fig:original_z1} closely follows the form \( A(t') \cos\left(2\pi t/\omega_r + \phi_1\right) \), with $\omega_r=\omega_0^2 + (3\lambda g^2/2\Omega^4)$. Consequently, increasing the fast forcing strength \( g \) effectively reduces the time period of the fast oscillations ($2\pi/\omega_r$), while, in contrast, it lengthens the time period of the slow envelope $A(t')$ modulating the amplitude. 
The impact of increasing $g$ in lowering the amplitude of \( \langle X \rangle \) and the fluctuation $V$ has been demonstrated in the following panels ($b,c$) ($e,f$). The parametric plot Fig.~\ref{fig:parametric_order} illustrates the interplay between the nonlinearity ($\lambda$) and the critical HFS-parameters ($g$, $\Omega$) in influencing the amplitude of the oscillation of $\langle X\rangle$. The regions of high and low amplitude oscillation are displayed in the corresponding panels ($a, b$, \text{and} $c$) for three distinct values of $\lambda: 0.001,0.01, \text{and}~ 0.1$. It is observed that the fast forcing mechanism suggested in this article functions fairly well for weak nonlinearity. Although increasing $g$ may not be an effective strategy for tuning fluctuations in the strongly nonlinear regime, it still exerts a degree of control over the amplitude. Another observation in \emph{Zone A} , where the effective natural frequency satisfies \( \omega_r = \omega_p/2 + \epsilon_r \delta \), clearly shows that raising \( g \) gradually causes the system to leave the resonance condition. Low-amplitude oscillations take over when \( \omega_r > \omega_p/2 + \epsilon_r \delta \).

For three distinct values of \( \lambda \), the approximate threshold values of \( g \) at which this transition is reflected have been determined. The threshold \( g \) values for the parameters \( \epsilon = 0.11 \), \( \delta = 0.125 \), \( \omega_0 = 0.5 \), and \( \omega_p = 1.0 \) are found to be roughly \(\{72, 17, 2\}\) for \(\Omega = 5\). Figure~\ref{fig:parametric_order} provides an illustration of these findings, showing the clear shift from high-amplitude to low-amplitude oscillations. Additionally, it is noted that when \( \lambda \) grows, the transition threshold also decreases.\\

\begin{figure}[h]
    \subcaptionbox{$g=0$}{\includegraphics[scale=0.425]{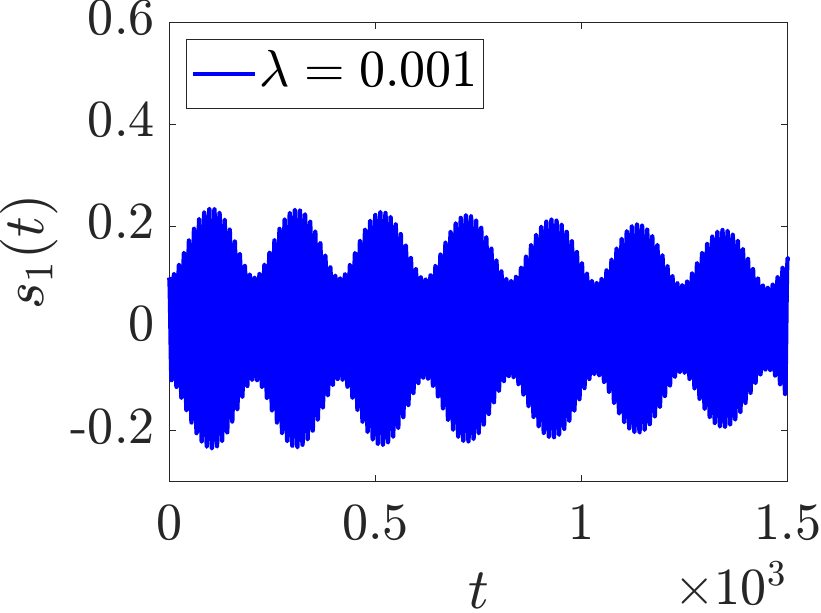}}
    \subcaptionbox{$g=60$}{\includegraphics[scale=0.425]{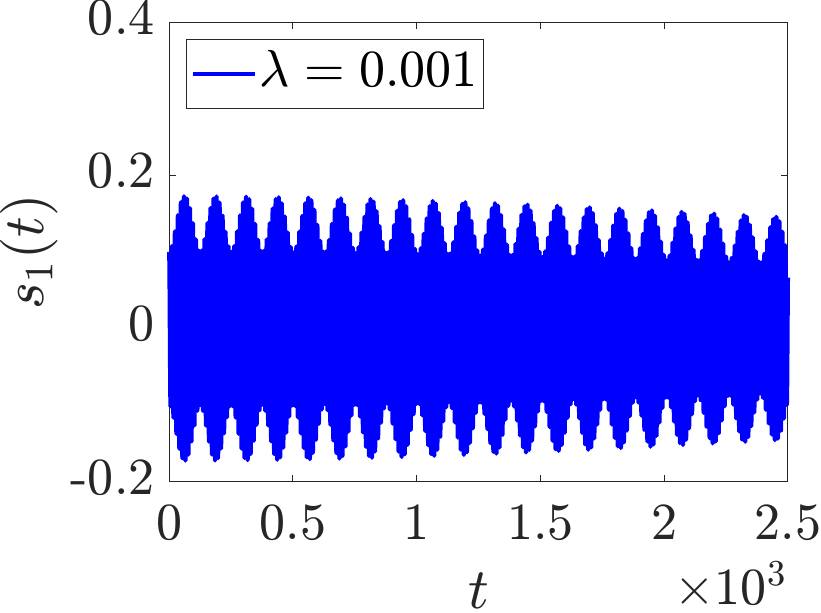}}
    \subcaptionbox{$g=100$}{\includegraphics[scale=0.425]{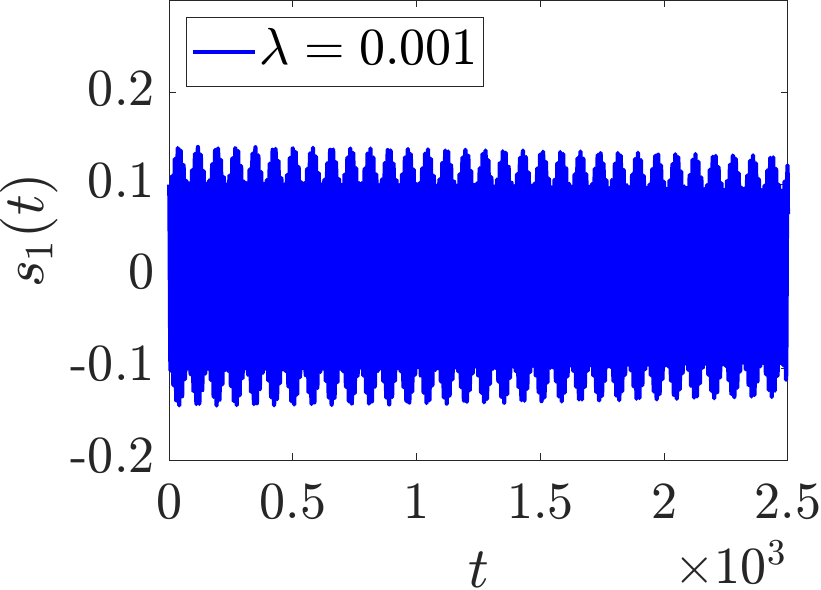}}
        
    \subcaptionbox{$g=0$}{\includegraphics[scale=0.425]{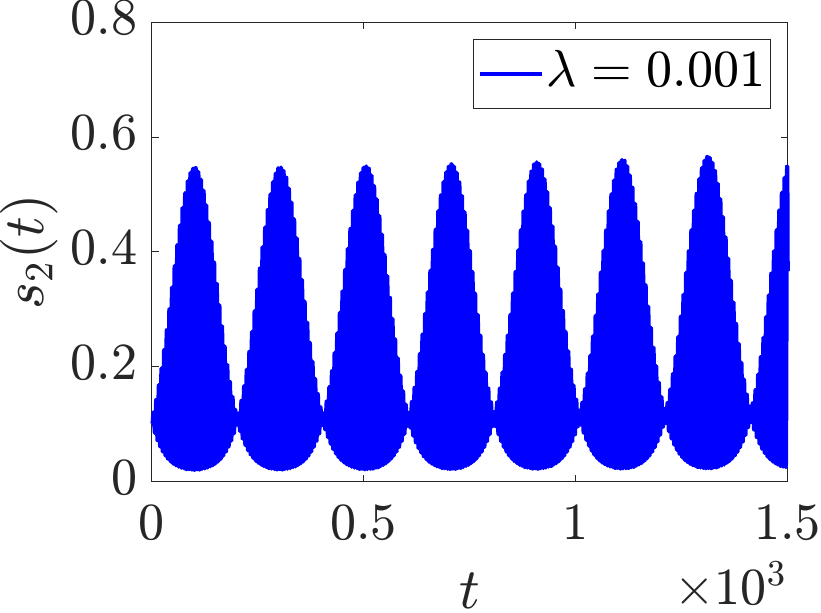}}
    \subcaptionbox{$g=60$}{\includegraphics[scale=0.425]{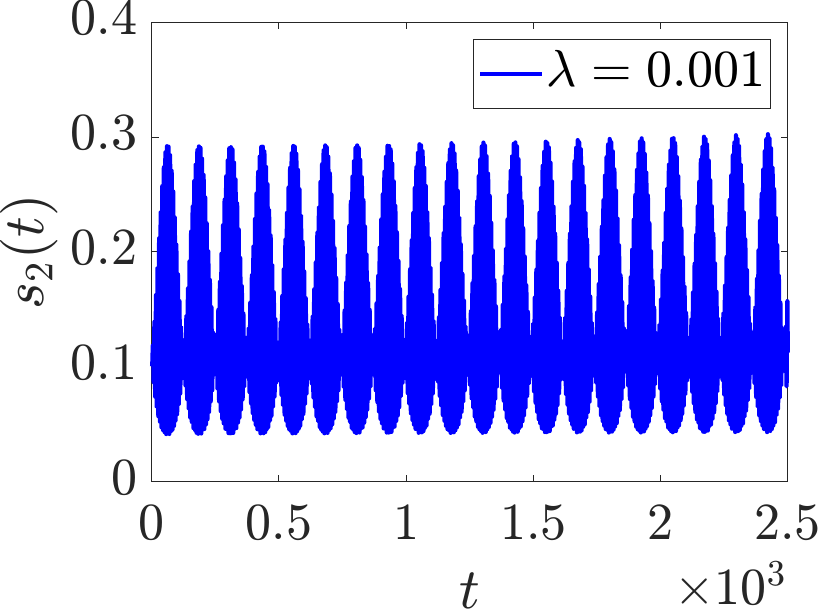}}
    \subcaptionbox{$g=100$}{\includegraphics[scale=0.425]{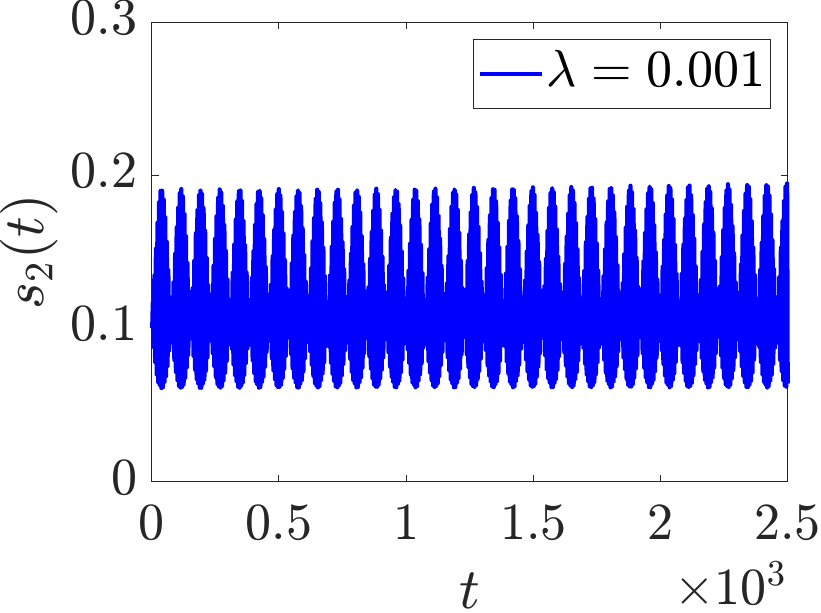}}
\caption{The original (effective) dynamics (\ie, Eq.~\eqref{Moment2}) in the \emph{Zone $B_2$} (\( \omega_0 > \omega_p/2 \)) are presented. The first row shows the evolution of the effective average \( \langle X(t) \rangle \) (or \( s_1 \)) for varying HFS strengths \( g = 0, 60, 100 \). The second row illustrates the corresponding effective dynamics of \( V(t) \) (or \( s_2 \)) under the same conditions. The other system parameters are \( \epsilon = 0.11 \), \( \omega_0 = 0.52 \),\( \omega_p = 1.0 \) and \( \Omega = 5.0 \).Dynamics shows the suppression of amplitude as $g$ is varied.
 The initial conditions used are \( s_1(0) = 0.1,\, \dot{s}_1(0) = 0 \) and \( s_2(0) = 0.1,\, \dot{s}_2(0) = 0,\, \ddot{s}_2(0) = 0 \).}
\label{fig:zoneB2}
\end{figure}

This shifting of \( \omega_r \) leads to a interesting phenomenon when \emph{Zone $B_1$} (\( \delta < -\omega_p/8 \)) is considered. As illustrated in Fig.~\ref{fig:schematic}, when \( \omega_r \) lies within this sub-zone, increasing \( g \) can push the system dynamics across the lower threshold of \emph{Zone A} (\( \omega_p/2 - \epsilon_r \delta \)) and into the resonance zone. The critical value of \( g \) at which the condition \( \omega_r > \omega_p/2 - \epsilon_r \delta \) is satisfied can be numerically estimated to be approximately 72, close to the point where resonant oscillations begin to appear. Upon further increasing \( g \), the system crosses the upper threshold of \emph{Zone A} (\( \omega_p/2 + \epsilon_r \delta \)) and enters \emph{Zone $B_2$}, resulting in low-amplitude oscillations. This upper threshold of \( g \) is found to be nearly 132. These arguments are clearly depicted in Fig.~\ref{fig:zoneB1} for \( \lambda = 0.001 \). A single value of \( \lambda \) is chosen for visual clarity; however, the same analysis can be applied for other \( \lambda \) values as well. Panel (a,d) shows the low-amplitude oscillations of \( \langle X \rangle \) and \( V \) in the absence of fast forcing, corresponding to \( \omega_r = \omega_0 = 0.47 < \omega_p/2 \). Panels (b,e) display the onset of resonant oscillations as \( g \) increases to 72, while panels (c,f) display the end of high-amplitude oscillations for $g=135$ and the transition of the system into \emph{Zone $B_2$}.
Finally, in Fig.~\ref{fig:zoneB2}, another regime (\emph{Zone $B_2$}) of low-amplitude periodic dynamics is portrayed. The behavior of this region closely resembles that of Zone $B_1$, except for the fact that if \( \omega_r \) is increased, the system remains periodic, maintaining low-amplitude oscillations until another subharmonic resonance zone appears. However, since this article primarily focuses on the primary parametric resonance (PSR) region (\( \omega_r \approx \omega_p/2 \)), the exploration of other resonance zones falls beyond the scope of this study.

\newpage
\section{Summary and Conclusions}
\label{sec6}
In this paper, the effect of fast external driving (HFS) on the rich dynamics of a quantum nonlinear parametric oscillator has been studied. The parametric and nonlinear nature of the oscillator, even at the classical level, give rise to interesting features like the presence of a series of subharmonic resonances and oscillation amplitudes that behave as the inverse of the nonlinearity parameter. The quantum version of this model includes the effect of higher order fluctuations (via a coupled hierarchy of equations) on the average dynamics of the oscillator and can be traced to the quantum anti-commutation relations of the canonical variables. Specifically, it has been shown that the introduction of a HFS can dramatically influence the above mentioned behavior. Since the equations governing the averages of the system variables are ultimately classical, a range of analytical techniques from classical nonlinear dynamics has been employed.

On one hand, it is found that the driven system can be treated via an averaging method which leads to an effective set of equations for the average of quantum variables; $\langle \hat{X} \rangle$ and $\langle (\hat{X}-\langle \hat{X} \rangle)^{2}\rangle$ (latter renamed $x(t)$ and $V(t)$ respectively in the paper)  have been the primary concern of the study. In fact, the strength and the frequency of the HFS conspire together to control the fixed points, amplitude of oscillations near the first subharmonic resonance, as well as other dynamical characteristics away from resonance. Secondly, since quantum mechanics dictates that the average of an operator is enslaved to its fluctuations ($V(t)$ in this case, but more generally to all higher orders in the presence of nonlinearity), this aspect helps in suppressing the high oscillation amplitudes around the resonance zones.

 An extensive analytical formulation of the problem has been carried out, and then the solutions have been explored at different parameter regimes via rigorous numerical techniques. Since the resonance zone is one of the focal points of the study, the linearisation method have been employed, slow flow equations have been derived and numerically solved to extract the physical implications of the critical fluctuation parameters and the influence of HFS. These results are then fully substantiated by direct numerical integration of the coupled effective equations of motion of $x(t)$ and $V(t)$ via which the claims about the effect/control via HFS are clearly shown.

Existing studies are complemented by the results of this paper as well as it points to several new directions of research. We have built on the previous works that employ the undriven linear and nonlinear parametric oscillators to probe the effect of quantum fluctuations via the classical nonlinear dynamical techniques \cite{sarkar2020nonlinear, biswas2019properties, biswas2021instability,sarkar2024quantum,pal2021quantum}. More importantly, to treat the HFS and derive the effective equations, the Bleckhman Perturbation Theory has been introduced, which analyzes the dynamics via splitting the fast and slow time scales of the system. In light of the method being recently used to study nonlinear Langevin equations dictated by stochastic forces \cite{DSRay1} and due to an easier route to numerical calculations, it is believed that a deeper understanding of fast/stochastically driven quantum oscillators can be gained and is a definitive future scope of work. Also,  the works on quantum parametric oscillators investigated by Dykman et al. have been complemented \cite{dykman2011quantum,zhang2017preparing,lin2015critical,marthaler2007quantum, boness2024resonant, marthaler2006switching}, and thus our work can motivate experimental realizations like in the case of qubits \cite{boness2025zero}.

Dissipation is the most important ingredient that has been left out in this work. On one hand, at the classical level it plays an important role in the dynamical characteristics of subharmonic resonances and stability/instability criteria in nonlinear parametrically driven systems. Quantum mechanically, a future direction of work is the analysis of the above-driven model (in various parameter regimes) coupled to a thermal bath. It is believed that this will have important physical implications for the control and modulation of such quantum devices in the presence of thermal fluctuations. Direct calculations of the rate of change of average energy of such a driven system covering a broad spectrum of driving strengths can be explored along with its connection to Thermodynamic Uncertainty Relations (TURs) \cite{horowitz2020thermodynamic} due to the natural emergence of higher order fluctuations.

Lastly, the hierarchy of moment equations that have been utilized for the driven quantum system may be improved by deferring the closure at higher ordered correlators. At the least, since the skewness and kurtosis of the distribution carry important information about the spreading of the initial wave-packet, their inclusion will lead to deeper physical insight about the quantum dynamics (for a system without nonlinearity and driving, see the analysis in Ref.~\cite{biswas2019properties}). This is in the spirit of of many other truncation methods that have been utilized and are still being investigated in a plethora of theoretical formulations to calculate correlators via hierarchical equations (in Quantum optics/Nonequilibrium/Disordered/Many-body systems, for example). Such an improvement will, of course, give rise to more analytical and computational complexities, but can also be complemented by exactly simulating the quantum Hamiltonian directly to test their validity. 

\newpage

\section{Acknowledgment}
SR sincerely expresses gratitude to Prof. Dr. Satyajit Chakrabarti (The honorable Director) for the opportunity to pursue a postdoctoral fellowship and to Prof. Dr. K.P. Ghatak (Emeritus professor) for his invaluable support, encouragement, and guidance. CB expresses his gratitude to the Department of Theoretical Physics at Universidad Complutense de Madrid for the opportunity to carry out research as a postdoctoral investigator. DB acknowledges research support from NBRC Gurgaon.

\appendix

\section{HEOM and Hierarchy of Moment Equations: Eqns.~\eqref{Moment0} and \eqref{Moment1}.}
\label{appendix:A}

\noindent The hierarchy equations are derived for the central moments ($\langle (\hat{X}-\langle\hat{X}\rangle)^{n}\rangle $) utilizing the Heisenberg Equations of Motion for the Hamiltonian:
\begin{eqnarray}
    \hat{H_{T}} &=&  \hat{H}_{S} + \hat{U} \nonumber\\
    \hat{H}_{S} &=& \frac{\hat{P}^{2}}{2} + \frac{1}{2}  \omega_{0} ^{2} (1+ \epsilon \cos{\Omega t}) \hat{X}^{2} + \frac{\lambda}{4} \hat{X}^{4} = \frac{\hat{P}^{2}}{2} + \hat{W}(\hat{X})\nonumber\\
    \hat{U} &=& G(t)\hat{X} = -g \hat{X}\cos{\Omega t} \label{H1.1}
\end{eqnarray}

\noindent The Heisenberg equation of motion (HEOM) for the expectation value of an operator is
\begin{eqnarray}
    \frac{d}{dt}\langle \hat{O} \rangle = \frac{\partial}{\partial t}\langle \hat{O} \rangle + \frac{1}{i \hbar} \langle[\hat{O},\hat{H}] \rangle \label{H3}
\end{eqnarray}

\noindent Ehrenfest's theorem naturally follow
\begin{eqnarray}
\frac{d}{dt} \langle \hat{X}\rangle &=& \langle \hat{P} \rangle \nonumber\\
\frac{d\langle \hat{P}\rangle }{dt} &=& -\langle \frac{d}{d\hat{X}} W(\hat{X})\rangle  \label {H4}
\end{eqnarray}

\noindent Let us briefly recap the undriven cases: $g=0$.

\noindent \textbf{Case 1: $g=\lambda=\epsilon=0$}:  If the Hamiltonian is atmost quadratic in powers of $\hat{X}$ (i.e. $\lambda =0$) and not parametrically driven ($\epsilon=0$), the above equations are closed. The dynamics is completely determined by the $\langle \hat{X}\rangle$ , $\langle \hat{P} \rangle$,$\langle \hat{X}^{2}\rangle$ , $\langle \hat{P}^{2} \rangle$, $\langle \hat {X} \hat{P} \rangle$ and $\langle  \hat{P}\hat {X} \rangle$.  It is intuitively obvious as in this case the wavefunctions/quantum states are Gaussian and the moments are relevant upto second order. All higher moments are zero for odd orders and products of the second moment if even. In other words, it is an exactly solvable problem.

\noindent \textbf{Case 2}: If $\lambda=g=0$ and $\epsilon \neq 0$, ie, the oscillator is parametrically driven, the hierarchy of moment equations are again closed at every order, and they represent Mathieu Equations (see Refs.~\cite{hanggi1993dissipative,biswas2019properties} for details.)

\noindent \textbf{Case 3} When, $g=0, \{\lambda, \epsilon\} \neq 0$, the above equations are not closed since in its most general form the second relation in Eq.~\eqref{H4} carries the signature of nonlinearity 
\begin{eqnarray}
    \frac{d}{dt} \langle \hat{P} \rangle &=& - \sum_{l=0}^{\infty} \frac{1}{l!}\langle \hat{X}^{l}\rangle \frac{d^{l+1}}{d^{l+1}\langle \hat{X}\rangle}  W(\langle\hat{X}\rangle) \label{H5} \
\end{eqnarray}
\noindent and an infinite hierarchy of moment equations is formed by the existence of all higher moments, which cannot be reduced to any finite set of lower-order moments. Also, mixed moments of the form $\langle \hat{X}^{n} \hat{P}^{m} \rangle$  come into play \cite{ballentine1998moment}.

\noindent \textbf{The model for $\hat{H}_{T}$:} The hierarchy is now explicitly derived for the model with $\{\lambda, \epsilon, g\} \neq 0$, using Eqs.~\eqref{H1.1}-\eqref{H4}. For the average, the following expression is obtained:
\begin{eqnarray}
    \frac{d^{2} \langle \hat{X} \rangle}{dt^{2}} = -\omega_{0}^{2} F(t) \langle \hat{X} \rangle- \lambda \langle \hat{X}^{3}\rangle + g\cos{\Omega t}\nonumber\\
    \implies \frac{d^{2} x}{dt^{2}} = -\omega_{0}^{2} F(t) x- \lambda x ^{3} -3\lambda V x -\lambda S + g\cos{\Omega t}
    \label{H5.1}
    \end{eqnarray} 
where, 
\begin{equation}
    \langle \hat{X} \rangle =x,\ V= \langle (\hat{X}-x)^{2}\rangle,\ \langle \hat{X}^{3} \rangle = x^{3} + 3Vx + S \label{H6},
\end{equation}
have been used.
\noindent Next, the same is calculated for the variance, 
$V=\langle \hat{X}^{2} \rangle-x^{2}$
\begin{eqnarray}
    \frac{d^{2} V}{dt^{2}}
    = 2(\Delta p)^{2} -2\omega_{0}^{2}F(t) V -2\lambda [ \langle \hat{X}^{4}\rangle- \langle \hat{X}^{3}\rangle x] \nonumber\\
    \implies  \frac{d^{2} V}{dt^{2}} = 2(\Delta p)^{2} -2\omega_{0}^{2}F(t) V -2\lambda [ K_{1} +3Vx^{2} -Sx] \label{H7}\nonumber\\ \label{H7}
    \end{eqnarray}
\noindent where, $(\Delta p)^{2} = \langle \hat{P}^{2} \rangle - \langle \hat{P} \rangle ^{2}$ and $\langle \hat{X}^{4} \rangle = x^{4} + 6V x^{2} + K_{1}$. The reader should note that this equation does not contain any $g-$dependent term as the FFT drives the oscillator linearly.
The final ingredient required is the time derivative of $(\Delta p)^2$, as provided by the HEOM.
\begin{eqnarray}
\frac{d(\Delta p)^{2}}{dt} =  -\omega_{0}^{2} F(t) \frac{dV}{dt} - \frac{\lambda}{2}\Bigg[\frac{d\langle \hat{X}^{4} \rangle}{dt} -4 \langle \hat{X}^{3} \rangle\frac{dx}{dt}\Bigg]\label{H8}
\end{eqnarray} 
\noindent Therefore, taking the time derivative of Eq.~\eqref{H7} and with a bit of algebra, we have
\begin{eqnarray}
    \frac{d^{3} V}{dt^{3}} +4\omega_{0}^{2}F(t)\frac{dV}{dt} + 2\omega_{0}^{2}\dot{F}(t)V &=&  -12 \lambda x^{2}\dot{V} - 12 \lambda V x \dot{x}\nonumber\\
    &&-3\lambda \frac{dK_{1}}{dt} + 6\lambda \frac{dS}{dt}x + 2\lambda S\frac{dx}{dt}\nonumber\\ \label{H9}
\end{eqnarray}

\noindent Equations~\eqref{H5.1} and \eqref{H9} make up the two equations in Eq.~\eqref{Moment0} in Sec.~\ref{sec2} of the main text with the identification
\begin{equation}
    \mathcal{F}(S,K_{1})\equiv -3\lambda \frac{dK_{1}}{dt} + 6\lambda \frac{dS}{dt}x + 2\lambda S\frac{dx}{dt}
\end{equation}

\noindent With the approximations $S = 0$ and $K = 3V^2$ applied to Eqns.~\eqref{H5.1} and \eqref{H9}, Eq.~\eqref{Moment1} in Sec.~\ref{sec2} of the main text is obtained.

\section{Derivation of effective dynamics: Eq.~\eqref{Moment2}.}
\label{appendix:B}
To derive the effective equations, Eq.~\eqref{MTA} is substituted into the first equation of Eq.~\eqref{Moment1}, leading to equations of the form
\begin{eqnarray}
\begin{split}
    \ddot{s_1}+\ddot{\psi_1}+\omega_0^2F(t)(s_1+\psi_1)
    +\lambda(s_1^3+3s_1^2\psi_1+3s_1 \psi_1^2+\psi_1^3)
    +3\lambda(s_1+\psi_1)(s_2+\psi_2)=g\cos(\Omega t).
\end{split}
\label{B1}
\end{eqnarray}
Taking the average of the above equation, while considering that the slow component remains nearly constant over the fast time scale, the following is obtained as
\begin{equation}
    \begin{split}
        \ddot{s_1}+\langle\ddot{\psi_1}\rangle +\omega_0^2F(t)(s_1+\langle\psi_1\rangle)
        +\lambda(s_1^3+3s_1^2\langle\psi_1\rangle+3s_1 \langle\psi_1^2\rangle+\langle\psi_1^3\rangle)
    +3\lambda(s_1+\langle\psi_1\rangle)(s_2+\langle\psi_2\rangle)=\langle g\cos(\Omega t)\rangle
    \end{split}
    \label{B2}
\end{equation}
By subtracting Eq.~\eqref{B2} from Eq.~\eqref{B1}, the following expression is obtained:
\begin{equation}
    \begin{split}
        \ddot{\psi_1}+3\lambda s_1(\psi_1^2-\langle\psi_1^2\rangle)+3\lambda(\psi_1^3-\langle\psi_1^3\rangle)
        +3\lambda(\psi_1\psi_2-\langle\psi_1\psi_2\rangle)=g\cos(\Omega t)-\langle g\cos(\Omega t)\rangle
    \end{split}
    \label{B3}
\end{equation}
The inertial approximation \cite{blekhman2000vibrational} assumes $\dddot{\psi}_j\gg\ddot{\psi}_j\gg\dot{\psi}_j\gg\psi_j,\psi_j^2,\psi_j^3...$, ${j=1,2}$ .So that Eq.~\eqref{B3} becomes
\begin{equation}
    \ddot{\psi_1} \approx g\cos(\Omega t),
    \label{B4}
\end{equation}
gives solution $\psi_1=-g\cos(\Omega t)/\Omega^2$ and the averages accordingly $\langle\psi_1^2\rangle=g^2/2\Omega^4, \langle\psi_1^3\rangle=0$. Upon substitution back into Eq.~\eqref{B2}, the effective slow dynamics for the variable $x$ is recast as
\begin{equation}
    \ddot{s_1}+\omega_r^2F_r(t)s_1+\lambda s_1^3+3\lambda s_1s_2=0
    \label{B5}
\end{equation}
Applying the same method, the second equation in Eq.~\eqref{Moment1} can be rewritten in the following form
\begin{equation}
    \begin{split}
        \dddot{s}_2+\dddot{\psi}_2+4\omega_0^2F(t)(\dot{s}_2+\dot{\psi}_2)+2\omega_0^2\dot{F}(t)(s_2+\psi_2)
        +12\lambda(s_1+\psi_1)^2(\dot{s}_2+\dot{\psi}_2)
        &+12\lambda(s_1+\psi_1)(s_2+\psi_2)(\dot{s}_1+\dot{\psi}_1)\\
        &+18\lambda(s_2+\psi_2)(\dot{s}_2+\dot{\psi}_2)=0
        \label{B6}
    \end{split}
\end{equation}
averaging over Eq.~\eqref{B6} gives

\begin{multline}
        \dddot{s}_2+\langle\dddot{\psi}_2\rangle+4\omega_0^2F(t)(\dot{s}_2+\langle\dot{\psi}_2\rangle)+2\omega_0^2\dot{F}(t)(s_2+\langle\psi_2\rangle)
        +12\lambda\bigg[s_1^2\dot{s}_2+s_1^2\langle\dot{\psi}_2\rangle+2s_1\dot{s}_2\langle\psi_1\rangle+2s_1\langle\psi_1\dot{\psi}_2\rangle\\
        +\dot{s}_2\langle\psi_1^2\rangle+\langle\dot{\psi}_2\psi_1^2\rangle\bigg]+12\lambda\bigg[s_1s_2\dot{s}_1+s_1\dot{s}_1\langle\psi_2\rangle+s_2\dot{s}_1\langle\psi_1\rangle
        +\dot{s}_1\langle\psi_1\psi_2\rangle+s_1s_2\langle\dot{\psi}_1\rangle+s_1\langle\dot{\psi}_1\psi_2\rangle+s_2\langle\dot{\psi}_1\psi_1\rangle+\langle\dot{\psi}_1\psi_1\psi_2\rangle\bigg]\\
        +18\lambda\bigg[s_2\dot{s_2}+s_2\langle\dot{\psi}_2\rangle+\dot{s}_2\langle\psi_2\rangle+\langle\psi_2\dot{\psi}_2\rangle\bigg]=0
        \label{B7}
\end{multline}
Now subtracting Eq.~\eqref{B7} from Eq.~\eqref{B6} yields the dynamics of the fast variable $\psi_2$ as:

\begin{multline}
        \dddot{\psi}_2 + 4\omega_0^2 F(t)(\dot{\psi}_2 - \langle\dot{\psi}_2\rangle) 
        + 2\omega_0^2 \dot{F}(t)(\psi_2 - \langle\psi_2\rangle) 
        + 12\lambda \bigg[ s_1^2 (\dot{\psi}_2 - \langle\dot{\psi}_2\rangle) 
        + 2s_1\dot{s}_2(\psi_1 - \langle\psi_1\rangle) 
        + 2s_1 (\psi_1\dot{\psi}_2 \\- \langle\psi_1\dot{\psi}_2\rangle) 
        + \dot{s}_2 (\psi_1^2 - \langle\psi_1^2\rangle) 
        + (\dot{\psi}_2\psi_1^2 - \langle\dot{\psi}_2\psi_1^2\rangle) \bigg]
        + 12\lambda \bigg[ s_1\dot{s}_1(\psi_2 - \langle\psi_2\rangle) 
        + s_2\dot{s}_1(\psi_1 - \langle\psi_1\rangle) 
        + \dot{s}_1(\psi_1\psi_2 \\- \langle\psi_1\psi_2\rangle)
        + s_1s_2(\dot{\psi}_1 - \langle\dot{\psi}_1\rangle) 
        + s_1(\dot{\psi}_1\psi_2 - \langle\dot{\psi}_1\psi_2\rangle)
        + s_2(\dot{\psi}_1\psi_1 - \langle\dot{\psi}_1\psi_1\rangle) 
        + (\dot{\psi}_1\psi_1\psi_2 - \langle\dot{\psi}_1\psi_1\psi_2\rangle) \bigg] \\
        + 18\lambda \bigg[ s_2(\dot{\psi}_2 - \langle\dot{\psi}_2\rangle) 
        + \dot{s}_2(\psi_2 - \langle\psi_2\rangle) 
        + (\psi_2\dot{\psi}_2 - \langle\psi_2\dot{\psi}_2\rangle) \bigg]= 0.
    \label{B8}
\end{multline}

It is important to note that $\psi_2$ implicitly depends on the scale $\psi_1$, which is evident since the fluctuation scale of $V$ varies with $x^2$ itself. The dominant contributions to Eq.~\eqref{B8} arise from $\psi_1^2\dot{\psi}_2$, $\psi_1\psi_2\dot{\psi}_1$ and $\psi_2\dot{\psi}_2$ . By incorporating the inertial approximation, as previously discussed, the effective equation for the system dynamics can be approximately obtained as
\begin{equation}
\dddot{\psi}_2+12\lambda\bigg[\psi_1^2\dot{\psi}_2+\psi_1\psi_2\dot{\psi}_1\bigg]+18\lambda\psi_2\dot{\psi}_2=0,
    \label{B9}
\end{equation}

Let $\dot{\psi}_2=z$, and using the value of $\psi_1$ and $\dot{\psi}_1$ the above equation yields
\begin{equation}
    \ddot{z}+\frac{6\lambda g^2}{\Omega^4}(1+\cos(2\Omega t))z-\frac{6\lambda g^2}{\Omega^3}zt\sin(2\Omega t)+8\lambda z^2 t=0,
    \label{B10}
\end{equation}
The above equation is highly nonlinear in nature and can be solved self-consistently. Given the smallness of $\lambda$, It is assumed that the predominant contribution comes from the term linear in $z$, which provides the zeroth-order solution as
\begin{equation}
    z_0=a_0\cos(\delta t+\eta),
    \label{B11}
\end{equation}
where $\delta=\sqrt{\frac{6\lambda g^2}{\Omega^4}}$. $a_0$ and $\eta$ are the amplitude and the phase, respectively, which are determined from the initial conditions.Thus, It can also be assumed that the dominant contribution to $\psi_2$ arises from the zeroth-order solution, though it can be extended to higher orders. However, for the sake of simplicity and without any loss of generality, we consider only the zeroth-order solution which gives 
\begin{equation}
    \dot{\psi}_2\approx z_0=a_0\cos(\delta t+\eta).
    \label{B12}
\end{equation}

It is clear that the dominant fast component of the fluctuation is also harmonic in nature as $\psi_2=-\frac{a_0}{\delta}\sin(\delta t+\eta)$.To this end, after determining the nature of \(\psi_1\) and \(\psi_2\), they can be directly substituted into Eq.~\eqref{B7}. Noting that \(\Omega \gg \delta\), all the averages in Eq.~\eqref{B7} over the scale \(\Omega\) approximately vanish, except for the average of \(\psi_1^2\). Consequently, the effective slow motion of the fluctuation is obtained as
\begin{eqnarray}
        \dddot{s}_{2} + 4\omega_{r}^{2} F_{r}(t) \dot{s_{2}} = -2 \omega_{r}^{2}\dot{F}_{r}(t)s_{2} - 12 \lambda s_{1}^{2} \dot{s}_{2}
- 18 \lambda \dot{s}_{2}s_{2} - 12 \lambda s_{2}\dot{s}_{1}s_{1},
 \label{B13}
\end{eqnarray}
Eqs.~\eqref{B5} and \eqref{B13} make up  Eq.~\eqref{Moment2} in the main text.

\bibliographystyle{apsrev4-1} 
\bibliography{citeDykman,citeJKB,cite_others} 

\end{document}